\documentclass[aps,prx,reprint,superscriptaddress]{revtex4-2}

\usepackage{amsmath}
\usepackage{amssymb}
\usepackage[utf8]{inputenc}
\usepackage{physics}
\usepackage{enumitem}
\usepackage{float}
\usepackage[margin = 2cm]{geometry}
\usepackage{color}
\usepackage{comment}
\usepackage{graphicx}
\usepackage{dsfont}
\usepackage{soul}
\usepackage{appendix}
\usepackage{braket}
\usepackage{chemformula}
\usepackage{dcolumn}
\usepackage{bm}
\usepackage{setspace}
\usepackage{xcolor}
\usepackage{booktabs}
\usepackage{tikz}

\usepackage{braket}
\usepackage{dcolumn}
\usepackage{bm}
\usepackage[colorlinks=true, citecolor=blue, linkcolor=blue, urlcolor=blue]{hyperref}
\usepackage{indentfirst}
\usepackage[dvipsnames]{xcolor}

\def \ETH{Institute for Theoretical Physics, ETH Z\"urich, CH-8093 Z\"urich, Switzerland}

\def \IBM{IBM Quantum, IBM Research – Zurich, 8803 R{\"u}schlikon, Switzerland}

\def \Rice{Department of Physics and Astronomy and Smalley-Curl Institute, Rice University, Houston, TX 77005, USA}

\begin{document}
\preprint{APS/123-QED}


\title{Programmable Polaron Models and Tomography in Trapped-Ion Quantum Simulators}


\author{Filip Marijanovi\'c}
\affiliation{\ETH}

\author{Utso Bhattacharya}
\affiliation{\ETH}
\affiliation{\IBM}

\author{Abhishek Menon}
\affiliation{\Rice}

\author{Visal So}
\altaffiliation[Present address: ]{California Institute of Technology, Pasadena, CA 91125}
\affiliation{\Rice}

\author{Guido Pagano}
\affiliation{\Rice}

\author{Eugene Demler}
\affiliation{\ETH}


\begin{abstract}
    Strong coupling between mobile charge carriers or excitons and lattice vibrations gives rise to polarons: quasiparticles formed by charge carriers dressed by a cloud of phonon excitations. Polaron formation determines transport properties of systems ranging from transition-metal oxides and organic semiconductors to light-harvesting complexes. In excitonic systems that admit a hard-core boson description, mapping the excitations to pseudospins provides a natural framework for studying polaron physics, including the nonperturbative regime in which hopping amplitudes, phonon energies, and coupling strengths are comparable. Here, we propose a trapped-ion blueprint for realizing spin-phonon Hamiltonians with programmable spatial coupling between excitation density and phonon displacements, encompassing local Holstein and long-range Fröhlich-type interactions. To access strongly dressed quasiparticles whose bare spectral weight is suppressed, we introduce displaced Green's functions implemented through spin-dependent phonon displacements. Their spectral response reveals the polaron band spectrum, including binding energies and phonon-excitation sidebands, while spatially resolved measurements probe the structure of the phonon dressing cloud, providing a route to polaron tomography. We further complement this spectroscopy with direct and ancilla-assisted measurements of spin-phonon correlations that track the coupled dynamics of the excitation and its environment. These approaches connect the emergent quasiparticle spectrum to the microscopic buildup of its dressing cloud, enabling controlled studies of how the spatial range of coupling shapes polaron formation and mobility.
\end{abstract}

\maketitle

\section{Introduction}
\label{sec:Introduction}

Electron--phonon interactions play a central role in determining the efficiency of exciton transport and charge-transfer processes in molecular aggregates, photosynthetic complexes, light-harvesting architectures~\cite{Fassioli2014, Scholes2017}, and polar semiconductors~\cite{Zhu2016screening, Miyata2017Large, Miyata2017lead, Franchini2021}. 
In solid-state systems, electron--phonon interactions govern emergent many-body phenomena in quantum materials, such as pairing in conventional superconductors~\cite{Bardeen1957} and the formation of charge-density waves~\cite{gruner2018density}, while their role in high-temperature superconductors remains a subject of considerable interest ~\cite{Egami2004role, Wang2021phonon, Tang2023traces, Yan2023ubiquitous}. 
Notably, even single-particle excitations can become strongly dressed by a phonon cloud at strong electron--phonon coupling, forming polarons~\cite{Landau1933, Holstein1959, Frohlich1954, AlexandrovDevreese}. This dressing results in modified carrier masses, quasiparticle spectral weights, transport coefficients, and optical properties~\cite{Franchini2021}. 

\begin{figure}[t]
    \centering
        \includegraphics[width=0.9\linewidth]{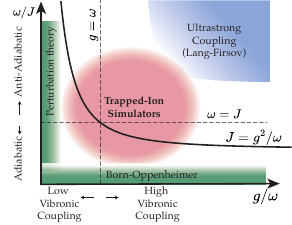}
    \caption{Qualitative diagram of the parameter regimes relevant to spin--phonon models of biomolecules \cite{Fassioli2014} and polarons \cite{Franchini2021}. Trapped-ion simulators naturally access the non-perturbative intermediate regime.}
\label{fig_regimes}
\end{figure}

Polaron physics is especially rich in regimes where the characteristic energy scales associated with electronic motion, lattice vibrations, and their mutual coupling become comparable (see Fig.~\ref{fig_regimes}). Such nonadiabatic regimes are increasingly relevant to a broad range of modern quantum materials, including transition-metal oxides~\cite{Mannella2005}, Moiré heterostructures~\cite{Sidler2017}, organic semiconductors~\cite{Coropceanu2007}, molecular aggregates~\cite{Spano2010}, and hybrid perovskites~\cite{Miyata2017Large}. The competition between electron hopping and self-trapping due to phonon dressing gives rise to rich dynamics that remain to be completely understood, as classical simulations of electron--phonon systems are severely constrained in these non-perturbative regimes. Despite decades of progress, existing approaches are typically effective only in specific cases: exact diagonalization~\cite{Fehske1999} is limited by system size and the rapid growth of the phonon Hilbert space; Monte Carlo methods~\cite{Prokofev1998, Berciu2006} are best suited to equilibrium properties and generally require analytic continuation to obtain energy spectra; tensor-network techniques~\cite{Jeckelmann1999DMRG} are most efficient for calculating ground-states \cite{Kohler2021} and linear response \cite{Jansen2023} in one dimension but become costly for long-range interactions~\cite{Frohlich1954} and nonequilibrium dynamics ~\cite{Gao2025}. Consequently, spectral functions, transport properties, and dynamical polaron formation across different coupling regimes remain difficult to access within a single classical framework, motivating their investigation using controllable quantum simulators.
Trapped ions are particularly well suited to studying lattice polaron physics~\cite{Mezzacapo2012, Stojanovic2012, Knoerzer2022} because they naturally arrange into self-assembled Coulomb crystals with well-defined equilibrium positions. This platform combines the individual control of highly coherent internal states and programmable long-range interactions~\cite{Porras2004, Monroe2021} with native, controllable bosonic modes and laser-induced spin--phonon couplings~\cite{Porras2008, Schneider_2012, Kang2024, pagano2025varenna}. 

In this work, we provide a detailed blueprint for engineering tunable spin–phonon Hamiltonians in which the spin-excitation density is controllably coupled to the displacement of phonon modes in multi-ion crystals and develop protocols to directly probe their equilibrium and nonequilibrium polaronic properties across different coupling regimes. 
Previous proposals for realizing local modes in polaronic Hamiltonians in trapped-ion systems have relied on microtraps and optical tweezers~\cite{porras2004bose, Stojanovic2012, gallagher2025quadratic, Knoerzer2022}.
Our architecture allows for continuous interpolation between Holstein-~\cite{Holstein1959, Holstein1959b} and Fr\"ohlich-type~\cite{Frohlich1954} regimes by directly controlling the spatial profile of the spin--phonon interactions~\cite{padilla2025vibrationallyassistedexcitontransfer}, as well as independent tuning of the long-range spin hopping, enabled by multi-tone site-resolved addressing and long motional mode coherence times~\cite{Feng2023, Than2025Yukawa, menon2026monolithic}.

Beyond providing an experimental pathway to realize tunable polaronic Hamiltonians, our work addresses a central challenge in quantum simulations of polaron physics: how to access ground-state and low-energy quasiparticle properties in strongly interacting regimes when direct ground-state preparation is challenging. We show that key equilibrium properties of polarons---including quasiparticle energies, effective-mass renormalization, binding energies, and transport characteristics---can instead be efficiently inferred from experimentally accessible nonequilibrium dynamical protocols. 

A key ingredient of our framework is the construction of observables that remain sensitive to strongly dressed quasiparticles. This is particularly important in the intermediate- to strong-coupling regime. Since conventional spectroscopic probes couple to bare spin operators, their overlap with dressed excitations becomes exponentially suppressed as the phonon dressing increases with spin--phonon coupling.
To overcome this limitation, we introduce displaced Green’s functions inspired by the Lang–Firsov transformation~\cite{LangFirsov1963} that can be measured in trapped-ion systems using spin-dependent displacements of phonons. By matching the measurement operator to the dressed quasiparticles, these correlators retain strong spectral weight for the low-energy polaronic excitations that become nearly invisible to conventional probes. Spatially resolved displaced local spectral functions further provide access to the structure of the dressing cloud, enabling a form of {\it polaron tomography}. 

Finally, we combine these observables into a comprehensive trapped-ion measurement toolbox that uses ancilla spins to probe quasiparticle and phonon dynamics simultaneously. Beyond displaced Green's functions, the toolbox provides access to spin--phonon correlations, phonon occupations, and phonon quadratures, allowing real-time tracking of polaron formation and the buildup of its dressing cloud.

Together, these measurements provide simultaneous access to the dressed quasiparticle and the microscopic dynamics of its phonon environment, enabling a level of characterization that is generally inaccessible in conventional condensed-matter experiments.

This paper is organized as follows. In Sec.~\ref{sec:Engineering_electron-phonon_models}, we introduce the tunable spin--phonon Hamiltonian in trapped-ion platforms, showing how independently driven orthogonal radial modes generate spin-exchange and spin--phonon interactions and how local phonon modes are constructed. In Sec.~\ref{sec:Holstein_and_Frohlich_intro}, we review the properties of Holstein and Fröhlich polarons and discuss polaron binding energies, effective mass renormalization, quasiparticle weight, and spectral functions. We extend our analysis to long-range hopping systems in  Sec.~\ref{sec:LDOS+Tomography}, introducing the local density of states (LDOS), the displaced local density of states (dLDOS), and \textit{polaron tomography}, which allows us to reconstruct the polaron wavefunction. In Sec.~\ref{sec:dynamical_mass_renorm}, we present a dynamical protocol using displaced initial states to extract effective mass renormalization from real-time evolution. In Sec.~\ref{sec:Long-range_hopping_polaron}, we validate the applicability of our protocols by simulating systems within realistic trapped-ion parameter ranges. In Sec.~\ref{sec:protocols}, we outline quantum measurement circuits utilizing ancilla ions to measure spin--phonon correlations and displaced Green’s functions in real-time. Finally, Sec.~\ref{sec:Discussion} provides an outlook toward multi-spin excitation regimes, bipolaron dynamics, higher-dimensional simulations, and simulations of doped $t-J$ models coupled to phonons, using qutrit systems.

\section{Tunable Spin-Phonon Models in Trapped-Ion Systems}
\label{sec:Engineering_electron-phonon_models}

\begin{figure*}[t!]
    \centering
    \includegraphics[width=1\linewidth]{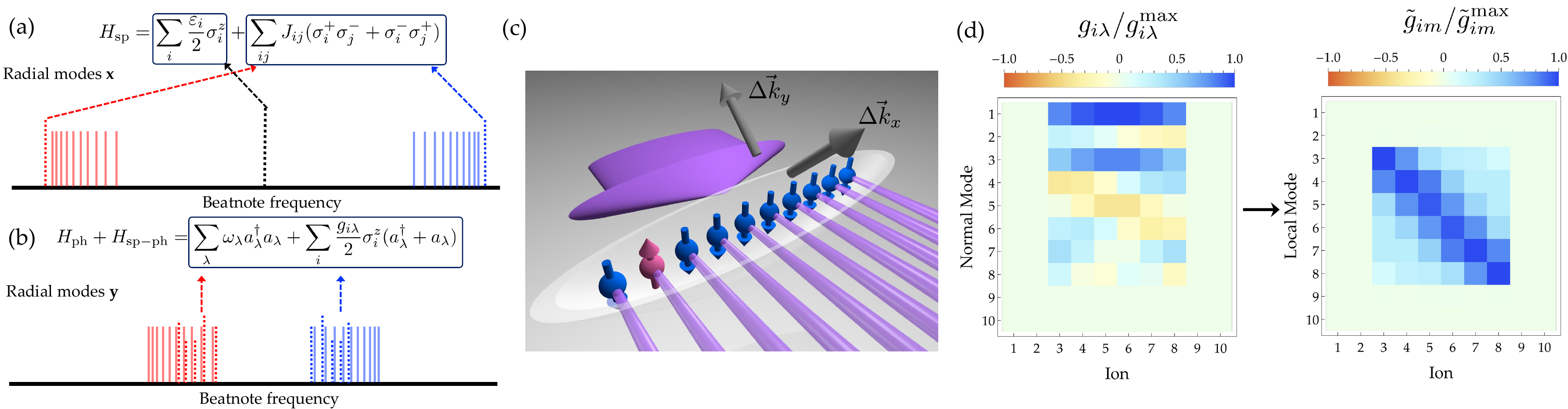}
    \caption{
\textbf{Trapped-ion architecture for programmable polaron models.}
(a) A far-detuned bichromatic drive couples to one radial motional branch along the ${\bf x}$ direction to generate
programmable spin--spin interactions and, in the presence of a strong transverse field, the effective spin-hopping Hamiltonian $H_{\rm sp}$.
(b) A near-resonant multi-frequency drive applied near the motional sidebands of the orthogonal radial branch along the ${\bf y}$ direction generates $H_{\rm ph}+H_{\rm sp-ph}$ with tunable effective phonon frequencies and spin--phonon coupling strengths $g_{i\lambda}$. 
(c) Optical geometry for independently coupling to the two radial directions. The laser wavevector $\Delta \vec{k}$ is decomposed along different radial principal axes of the trap potential.
(d) Optimized $g_{i\lambda}$ matrix describing the coupling between spin site $i$ and normal mode $\lambda$, designed using the RLP mapping, to realize a target local-mode matrix $\tilde{g}_{im}$ corresponding to a Fr\"ohlich Hamiltonian with $\alpha=1.6$ for $N=10$ ions. The optical drive is optimized to use only $M=8$ normal modes, driven on the $N_s=6$ central ions, leaving four uncoupled ancillas for measurement purposes (see Sec. \ref{sec:protocols}).   
}
\label{fig:architecture}
\end{figure*}

Trapped-ion crystals naturally combine long-lived internal electronic states with quantized collective motion. Two internal states of each ion encode an effective spin-$1/2$ degree of freedom, while the normal modes of the crystal provide bosonic degrees of freedom. Laser-induced optical dipole forces can be operated in different detuning regimes {with respect to the motional mode frequencies} to generate either effective spin--spin interactions, for which the motion is only virtually excited, or near-resonant spin--phonon interactions, for which selected motional modes participate as dynamical degrees of freedom

\cite{Porras2004, Porras2008, Schneider_2012, Monroe2021, so2024trappedion, padilla2025vibrationallyassistedexcitontransfer, so2025, katz2023programboson, pagano2025varenna}.
This combination makes trapped ions particularly well suited for simulating paradigmatic electron--phonon models: spin excitations represent itinerant particles whose hopping is generated by spin--spin exchange, while the crystal vibrations represent the lattice phonons that dress their motion. {The general form of the spin--phonon Hamiltonian considered in this work is
\begin{equation}
    \begin{split}
    H =& \underbrace{\sum_i \frac{\varepsilon_i}{2} \sigma^z_i +\sum_{i<j} J_{ij} (\sigma_i^+ \sigma_j^- + \sigma^-_i\sigma^+_j )}_{H_{\rm sp}} + \\ &  \underbrace{\sum_{\lambda} \omega_{\lambda} a^\dagger_{\lambda} a_{\lambda}}_{H_{\rm ph}} +  \underbrace{\sum_{i\lambda} g_{i\lambda} n_i  (a^\dagger_{\lambda} + a_{\lambda})}_{H_{\rm sp-ph}}.
    \label{eq_tot_H_ions}
    \end{split}
\end{equation}
The first term, $H_{\rm sp}$, describes the effective electronic sector, in which a spin excitation hops between sites with amplitude $J_{ij}$ and experiences an on-site potential $\varepsilon_i$. The second term, $H_{\rm ph}$, describes the phonon modes. The final term, $H_{\rm sp-ph}$, couples the local occupation of the spin excitation to the displacement of these modes, where the excitation-number operator at site $i$ is $n_i=\sigma_i^+\sigma_i^-=\left(1+\sigma_i^z\right)/2$. Accordingly, the spin--phonon interaction can equivalently be written as
\begin{equation}
    H_{\rm sp-ph}=\sum_{i\lambda} \frac{g_{i\lambda}}{2} \sigma^z_i  (a^\dagger_{\lambda} + a_{\lambda})
    +
    \sum_{i\lambda}\frac{g_{i\lambda}}{2}
    \left(a_\lambda^\dagger+a_\lambda\right).
    \label{eq:sigmaz_number_equivalence}
\end{equation}
}

The central idea of our scheme for realizing the tunable-range spin--phonon model in Eq.~\eqref{eq_tot_H_ions} is illustrated in
Fig.~\ref{fig:architecture}. We use the two orthogonal radial branches of the ion crystal for distinct purposes. One branch
mediates the effective hopping of spin excitations. A bichromatic laser field is detuned far from the sidebands of this branch, such that its motional modes are only virtually populated. Adiabatic elimination of these modes produces a programmable long-range Ising interaction. In the presence of a strong global field perpendicular to the Ising-interaction axis, terms that create or annihilate pairs of spin excitations rapidly average to zero, leaving the excitation-number-conserving exchange Hamiltonian
\begin{equation}
    H_{\rm hop}
    =
    \sum_{i<j}J_{ij}
    \left(
    \sigma_i^+\sigma_j^-+\sigma_i^-\sigma_j^+
    \right).
\end{equation}
Here, $\sigma_i^{+,-,z}$ are the Pauli operators acting on site $i$, and $\sigma^{\pm}_i = (\sigma^x_i \pm i \sigma^y_i)/2$ {are the corresponding spin raising and lowering operators}. The overall strength and spatial range of $J_{ij}$ are controlled by the laser intensities and detunings. Over experimentally relevant parameter regimes, the couplings are approximately described by $J_{ij}\simeq J/d_{ij}^{\alpha_J}$, where $d_{ij}\equiv|i-j|$ is the distance between sites $i$ and $j$, and the interaction-range exponent can typically be tuned within $0.5<\alpha_J<1.8$.

The other orthogonal radial branch, whose collective modes are described by operators
$a_\lambda$ and $a_\lambda^\dagger$, is instead retained as the dynamical phonon sector. As shown in
Fig.~\ref{fig:architecture}(b), a multi-frequency drive is applied close to the sidebands of selected modes of this branch. In this near-resonant regime, the detuning of $-\omega_\lambda$ of the tone addressing normal mode $\lambda$ sets its effective phonon energy to $\omega_\lambda$ in the rotating frame~\cite{Schneider_2012,so2024trappedion,pagano2025varenna}. The amplitudes and phases of the individual tones determine the spin--phonon coupling matrix $g_{i\lambda}$ and the phonon quadrature coupled to each spin. An appropriate choice of the spin phase and interaction basis~\cite{pagano2025varenna} produces a longitudinal coupling proportional to
$\sigma_i^z\,x_{\lambda} $ where the displacement is defined as $x_\lambda=a^\dagger_\lambda+a_\lambda$, corresponding to the first term in Eq.~\eqref{eq:sigmaz_number_equivalence}.

The spin-independent linear phonon term in Eq.~\eqref{eq:sigmaz_number_equivalence} can be eliminated in two ways. The first method is to directly implement this interaction. The second method is by a displacement transformation that shifts each mode to its displaced equilibrium position in the rotating frame. This transformation leaves the spin--phonon interaction unchanged in form, while renormalizing the on-site energies and introducing an overall constant energy shift (see Appendix~\ref{sec:app-deriving_Hamiltonian}). Since both methods require considerable displacements in the order of $\sim \sum_i g_{i\lambda}$ per phonon mode $\lambda$, they can be efficiently implemented with additional spin-independent forces on the phonons using time-varying electric fields, phase-coherent with the other multi-frequency drives, applied on the ion-trap's electrodes \cite{home2011normalinanharmonic}. The spectral separation of the two collective motional branches allows $J_{ij}$, $g_{i\lambda}$, and $\omega_\lambda$ to be controlled largely independently. Residual off-resonant spin--spin couplings generated by the multi-frequency spin--phonon drive are discussed in Appendix~\ref{sec:app-deriving_Hamiltonian}. Finally, local carrier drives can provide independently programmable on-site energy shifts $\varepsilon_i$. The accessible parameter ranges are summarized in Table~\ref{table:param_range}. 


As shown in the next section, transforming from the collective normal phonon mode basis to the local phonon basis is useful to reveal the structure of the phonon dressing cloud around an itinerant spin excitation.
We introduce the L\"owdin orthogonalization \cite{lowdin1950orthogonalization} as the general formalism to map normal phonon modes to local phonon modes. The transformation follows
\begin{equation}
    g_{i\lambda}=\sum^{N_{\rm s}}_{m} \tilde{g}_{im} V_{m\lambda}\,,
    \label{eq:lowdin_transform_g}
\end{equation}
where $\tilde{g}_{im}$ is the target spin-local-phonon coupling between ion $i$ and local mode $m$ (where $N_{\rm s}$ is the number of ions addressed) using the spin-phonon drive $g_{i\lambda}$. $V_{m\lambda}$ is the L\"owdin-orthogonalized transformation matrix relating operators in the normal mode basis $\{\lambda\}$, to the local phonons of mode basis $\{m\}$, and follows $\sum^M_\lambda V_{m\lambda} V_{\lambda m{'}}= \delta_{m m{'}} $, where $M$ is the number of addressed normal modes and $N$ is the total number of ions. 
Eq.~\eqref{eq:lowdin_transform_g} implies that a unique, exact multi-tone spin-phonon drive $g_{i\lambda}$ can be applied to achieve the desired $\tilde{g}_{im}$.
The local mode annihilation operator $a_m$ can be expressed in terms of the normal mode operators as
\begin{equation}
    a_m=\sum^M_{\lambda}V_{m\lambda}a_\lambda\,.
    \label{eq:lowdin_operator_transform}
\end{equation}
Furthermore, $V_{m\lambda}$ is a gauge matrix of dimension $N_{\rm s} \times M$, which means it can fulfill $VV^T=\mathbb{I}_{N_{\rm s}}$ while satisfying other constraints (see Supplemental Material).

We discuss three methods of local phonon mapping under this general formalism, distinguished by the extent to which phonon locality in real space is preserved:

{\bf\emph{(i)}} \emph{Drive-induced real-space local phonons} (RLP): We obtain the drive-induced real-space local phonon operators, such as annihilation operator $a_m$, by using the transformation matrix $V_{m\lambda}=\xi_{m\lambda}$ (size $N\times N$), which is the normal-mode eigenvector matrix mapping the collective mode basis $\{\lambda\}$ strictly to the real-space local mode (or site-dependent) basis $\{m\}$.
The spin--phonon couplings then transform as $\tilde{g}_{im}=\sum^N_{\lambda} g_{i\lambda}\xi_{\lambda m}$.
The solution is exact for a general $\tilde{g}_{im}$ only when all the normal modes are driven, i.e., $M=N$ non-zero columns exist in $g_{i\lambda}$. 
In the case of $M<N$, an approximate solution to $\tilde{g}_{im}$ can always be obtained by determining a $g_{i\lambda}$ matrix from numerically minimizing an appropriate cost function, such as $(||\tilde{g}^{\rm target}_{im}-\tilde{g}_{im}||_2)^2$ (see Fig.~\ref{fig:architecture}(d) and Supplemental Material).

{\bf\emph{(ii)}} \emph{Wannier-like local phonons} (WLP):
The approximate solutions of $\tilde{g}_{im}$ in the RLP case can be modified to be exact by defining instead a quasi-local Wannier-like phonon mode basis \cite{Cockayne2005}, which maximizes real-space locality constrained by the number of normal modes participating in the spin-phonon multi-frequency drive matrix $g_{i\lambda}$ (size $N_{\rm s}\times M$). Here, $V_{m\lambda}\neq\xi_{m\lambda}$, but $V_{m\lambda}$ (size $N_{\rm s} \times M$) can be tuned to maximally overlap with the locality defined by $\xi_{m\lambda}$,
while still providing an orthogonal set of $N_{\rm s}$ quasi-local modes (see Supplemental Material). Alternatively, $V_{m\lambda}$ can be optimized for minimizing the total laser power and/or minimizing the residual off-resonant spin–spin couplings induced by $g_{i\lambda}$.  The exact solution for $\tilde{g}_{im}$ exists for $N_{\rm s}\leq M\leq N$, but approximate solutions can be obtained for $M<N_{\rm s}$ using a numerical optimizer similar to the RLP method.

{\bf\emph{(iii)}} \emph{Synthetic local phonons} (SLP): The local modes in spin-phonon models can be directly mapped to the native normal modes of the trapped-ion chain, defining a synthetic-space lattice at the expense of real-space locality. Independent normal phonon modes are non-local in real space, but their coupling to the spin can be enforced to maintain locality in their synthetic space~\cite{tranpowerlawLRbound2019, junemannspinbosonLRbound2013} by encoding a decaying spin-phonon interaction graph $\tilde{g}_{im}$ which is the strongest for $i=m$. Here, $V_{m\lambda}= \mathbb{I}_{m\lambda}$ (size $N_{\rm s} \times N_{\rm s}$), and local mode $m\in\text{Perm}(\lambda)$, which is a suitable permutation of the normal mode set $\{\lambda\}$. Crucially, this simplifies the experimental complexity in:
{\bf\textit{(a)}} engineering the spin-phonon interaction graph to $\tilde{g}^{\text{target}}_{im}=g_{i\,\text{Perm}(\lambda)}$, 
{\bf\textit{(b)}} simulating highly local Holstein models accurately,
{\bf\textit{(c)}} implementing polaron tomography (see Sec.~\ref{sec:LDOS+Tomography}), and
{\bf\textit{(d)}} realizing dispersive phonons without phonon-phonon hopping terms
(see Supplemental Material).

We note that our protocols for polaron tomography in Sec.~\ref{sec:LDOS+Tomography} are agnostic to the above phonon mapping methods.

\begin{table}[h!]
\centering
\begin{tabular}{lll}
\toprule
\textbf{System parameter} & \medspace \textbf{Experimental range} \\
\text{($\hbar = 1$)} & \medspace \text{$[2\pi \times ~\text{kHz}]$}\\
\midrule
On-site energy shift $|\varepsilon_i|$ & \medspace $\leq20 $ \\ 
Spin-spin & \medspace $\leq0.6$, $0.5\lesssim \alpha_J \lesssim 1.8$ \\
coupling $J_{ij}=J/d_{ij}^{\alpha_J}$\\
Phonon energy $\omega_{\lambda}$ & \medspace $\sim0.5\text{--}10$\\ 
Spin-phonon coupling $g_{i\lambda}$ & \medspace $\sim0.5\text{--}3$\\ 
\bottomrule
\end{tabular}
\caption{Experimental range of system parameters used in the simulations of Sec.~\ref{sec:dynamical_mass_renorm}. Here, the maximum attainable carrier Rabi frequency per tone $\Omega^{\rm max}$ is capped to $2\pi\times$155 kHz. The middle 6 ions of a 10-ion chain are addressed with a waveform generating the Hamiltonian in Eq.~\eqref{eq_tot_H_ions}, where $M=8$ normal modes are coupled to the spins.}
\label{table:param_range}
\end{table}

\section{Holstein and Fr\"ohlich polarons}
\label{sec:Holstein_and_Frohlich_intro}

In this work, we study the general Hamiltonian in Eq.~\eqref{eq_tot_H_ions} within the single-excitation sector, consisting of one spin-up excitation on top of a spin-down background. We focus on the case of dispersionless (optical) phonon modes with $\omega_\lambda = \omega$, most relevant for solid-state systems,
and operate in the regime $\omega \gtrsim J $ (see Fig.~\ref{fig_regimes}), where the polaron effects become important.
In this limit, phonon modes are degenerate, and the phonon Hamiltonian is invariant under a change from the collective normal-mode basis to a spatially local-mode basis. Furthermore, the L\"owdin orthogonalization-based mapping linearly transforms $g_{i\lambda} \rightarrow \tilde g_{im}$, where $m$ is a real-space or synthetic-space index. 

We consider two representative hopping regimes. The first is nearest-neighbor (NN) hopping with $J_{ij} = J \delta_{i,j\pm 1}$, which corresponds to the tight-binding limit typically arising in solid-state systems. The second is power-law hopping with $J_{ij} = J/d_{ij}^{\alpha_J}$ and $\alpha_J = 1.2$, representative of the long-range interactions naturally realized in trapped-ion platforms~\cite{Monroe2021,pagano2025varenna} (see Table~\ref{table:param_range}). Extensions incorporating long-range hopping have become increasingly important for describing polaron formation and charge transport in correlated and low-dimensional materials. For example, in several models of cuprates and related correlated systems, beyond-nearest-neighbor hopping and electron-phonon coupling are believed to play an important role in shaping the quasiparticle dispersion and the superconducting transition \cite{Devreese2009, Wang2021phonon, Khatami2008}. Studies of generalized Holstein models with next-nearest-neighbor hopping similarly show significant modifications of polaron formation and transport \cite{Chandler2016, Chakraborty2011, Chakraborty2023}.

Minimal lattice models capturing the polaronic phenomena are typically formulated in terms of the generalized Holstein- or Fr\"ohlich-type Hamiltonians.  The generalized Holstein model describes short-range spin-phonon coupling given by
\begin{equation}
    \tilde g_{im}^{\text{H}} = \tilde {g} e^{-|i-m|/\xi},
\end{equation}
where $\xi$ is the range of interaction, which can be experimentally tuned. In the limit of $\xi \rightarrow 0$, it corresponds to the \textit{on-site} Holstein model, in which the spin at site $i$ couples only to the local lattice displacement, $\tilde g_{im}^{\text{H}} = \tilde {g} \delta_{i,m}$. The Holstein model serves as a simple model \cite{Holstein1959, Holstein1959b}, for short-range electron--phonon interactions in molecular crystals, organic semiconductors, and correlated lattices, where screening effects reduce the range of Coulomb interaction. 
By contrast, in systems with weak screening, such as ionic crystals, polar semiconductors, and trapped charge carriers in dielectric environments, the spin-phonon interactions remain long-range. The spin-phonon coupling is then given by the Fr\"ohlich model defined by~\cite{Frohlich1954,PhysRevB.61.12315}   
\begin{equation}
    \tilde g_{im}^{\text{F}} = \frac{\tilde {g}}{(|i-m|^2 + 1)^{\alpha/2}}.
\end{equation} 
Here, $\alpha$ controls the spatial decay of the spin--phonon interactions. More broadly, Fr\"ohlich-type interactions produce qualitatively distinct dressing compared to local Holstein limits \cite{Alexandrov2003}, emphasizing the need for platforms capable of realizing both short- and long-range electron--phonon couplings. 

We study the Hamiltonian in Eq. \eqref{eq_tot_H_ions} using exact diagonalization, for a system of $6$ spin sites with $6$ different phonon modes and a phonon number cutoff of $6$ quanta 
per mode. The lowest energy scale in our system is defined to be the nearest-neighbor hopping strength $J$. In the case of `Trapped-Ion' mode simulations (see Sec. \ref{sec:Long-range_hopping_polaron}), we focus on the case of $8$ bosonic modes and reduce the phonon number cutoff to $4-5$ quanta per mode (see Fig. \ref{fig:long-range_hol_vs_fro} for details). We construct the Hilbert space around a displaced coherent state~\cite{Bhattacharya2021PhononInducedPairing, Zhang2023PeierlsNanotube}, allowing us to study the model in the strong coupling regime, without requiring large phonon cutoffs. Using the Lanczos algorithm, we find the relevant Hilbert subspace for diagonalizing the Hamiltonian. We supplement our results with analytics based on Lang-Firsov transformation (see Sec.~\ref{sec:LF}), which are valid in the strong-coupling limit.

\subsection{Properties of Holstein and Fr\"ohlich models} \label{sec:LF}

We now review the known properties and signatures of the polaron formation in the Holstein and Fr\"ohlich models with NN hopping. Our goal is to set the stage for subsequent discussion of quantum simulations of polarons with ion chains. The key signatures of polaron physics are the formation of the polaron cloud, a strong reduction in the quasiparticle residue, and an exponential increase in the effective mass of the dispersive polaron with coupling strength.
In addition, the excitation spectrum consists of single-phonon excitations on top of the polaron ground state~\cite{AlexandrovDevreese}. To understand these properties, we study the Hamiltonian in Eq.~\eqref{eq_tot_H_ions} using the Lang-Firsov transformation (LF)~\cite{LangFirsov1963} to rotate the Hamiltonian into the polaron frame, where the particles are dressed by phonon clouds, and the interaction is encoded in renormalized hopping amplitudes and energy shifts. The transformation is generated by the unitary
\begin{equation}
    U_{\rm LF} = \exp\left(\sum_{i,\lambda} \frac{g_{i\lambda}}{\omega_\lambda} n_ia_{\lambda} - \text{h.c.}\right).
    \label{eq:langfirsov}
\end{equation}
By assuming phonons are in vacuum (for details, see Appendix \ref{sec:app-Lang-firsov}), we obtain the effective spin Hamiltonian within the single-spin sector
\begin{equation}
    H^{\rm eff} = \sum_{i < j} J_{ij}^{\rm eff} (\sigma_i^+ \sigma_j^- + {\rm h.c.}) - \sum_{i, \lambda} \frac{\abs{g_{i\lambda}}^2}{\omega_\lambda} n_i,
    \label{eq:eff_hamiltonian}
\end{equation}
where the effective hopping is renormalized by the phonon cloud overlap
\begin{equation}
    J_{ij}^{\rm eff} = J_{ij} \exp\left( - \frac{1}{2} \sum_\lambda \left|\frac{{g}_{i\lambda} - {g}_{j\lambda}}{\omega_\lambda}\right|^2\right).
    \label{eq:eff_hopping}
\end{equation}
Therefore, the presence of phonons has two main effects on the system. First, they create an effective potential for the spins, as described by the second term in Eq.~\eqref{eq:eff_hamiltonian}, and second, they renormalize the hopping coefficients in the spin chain, as shown in Eq.~\eqref{eq:eff_hopping}. For the case of local phonon modes the above equation can be simply rewritten by replacing $\omega_\lambda \rightarrow \omega_m$ and $g_{i\lambda} \rightarrow \tilde{g}_{im}$. We note an important difference between periodic translationally invariant and finite open boundary systems: in a translationally invariant system, the coupling $\tilde{g}_{im}$ only depends on the distance $|i-m|$, and, therefore, the effective potential becomes $-\sum_{i, m} \left(\tilde {g}^2_{i-m}/\omega_m\right)n_i$. For periodic systems, the sum over $m$ results in a constant contribution, independent of spin position $i$. For finite systems, however, the sum gives rise to a finite local potential,  except in the case of on-site Holstein coupling with $\tilde {g}_{i-m} = \tilde {g}\delta_{i,m}$.

\subsection{Spectral function}

To understand the effects of polaron dressing, we study the spectral function of the system because it encodes both the energy spectrum of the system and the quasiparticle weight, which quantifies how strongly the original spin excitation overlaps with a given quasiparticle mode. A large weight indicates a well-defined, particle-like excitation, while a smaller weight signals strong interactions or dressing by the environment. To obtain the spectral function, the key idea is to measure the Green's function $G(j,l,t)$, which gives the amplitude for a spin excitation created at site $j$ on the vacuum state $\ket{0_{\rm ph}}$ to be found at site $l$ after a time $t$,
\begin{equation}
    G(j,l, t) = -i \theta(t)\bra{l} e^{-iHt} \ket{j},
\end{equation}
where $\ket{j} = \sigma_j^+ \ket{\downarrow}_N \otimes \ket{0_{\rm ph}}$ represents a single (bare) spin excitation at site $j$ with all phonons in vacuum, $\ket{\downarrow}_N = \Pi_i \ket{\downarrow}_i$ and $\theta(t)$ is the Heaviside theta function. The experimental protocol for measuring $G(j,l,t)$ based on a Hadamard test \cite{Somma2004, DelRe2024} is detailed in Sec.~\ref{sec:protocols}. 

We now focus on the case of the on-site Holstein model, where at strong coupling, there is no additional potential generated by the phonons (see Sec.~\ref{sec:LF}). 
In this case, the total quasi-momentum $k$ is still a good quantum number (see Appendix \ref{sec:app-obc} for details) since the system possesses open boundary conditions (OBC) and only NN hopping. To compute the spectral function, we spatially transform the Green's function using the OBC eigenfunctions and perform a temporal Fourier transform, giving $A(k,\nu) = - \frac{1}{\pi} \Im{G(k,\nu)} $. In terms of the many-body eigenstates $\ket{\psi_n(k)}$ with energy $E_n(k)$, the spectral function can be expressed as
\begin{equation}
    A(k,\nu) = \sum_n |\langle \psi_n(k) \ket{k,0_{\rm ph}}|^2 \delta(\nu - E_{n}(k)),
    \label{eq:spectral_function}
\end{equation}
where $\ket{k,0_{\rm ph}}$ is the quasi-momentum $k$ state. The above equation, explicitly shows that the spectral function contains both the full spectral information of the system encoded in $\delta(\nu - E_{n}(k))$, as well as the quasiparticle weights of the eigenstates, encoded in the matrix elements $|\langle \psi_n(k) \ket{k,0_{\rm ph}}|^2$. 
\begin{figure}
    \centering
    \includegraphics[width=1.0\linewidth]{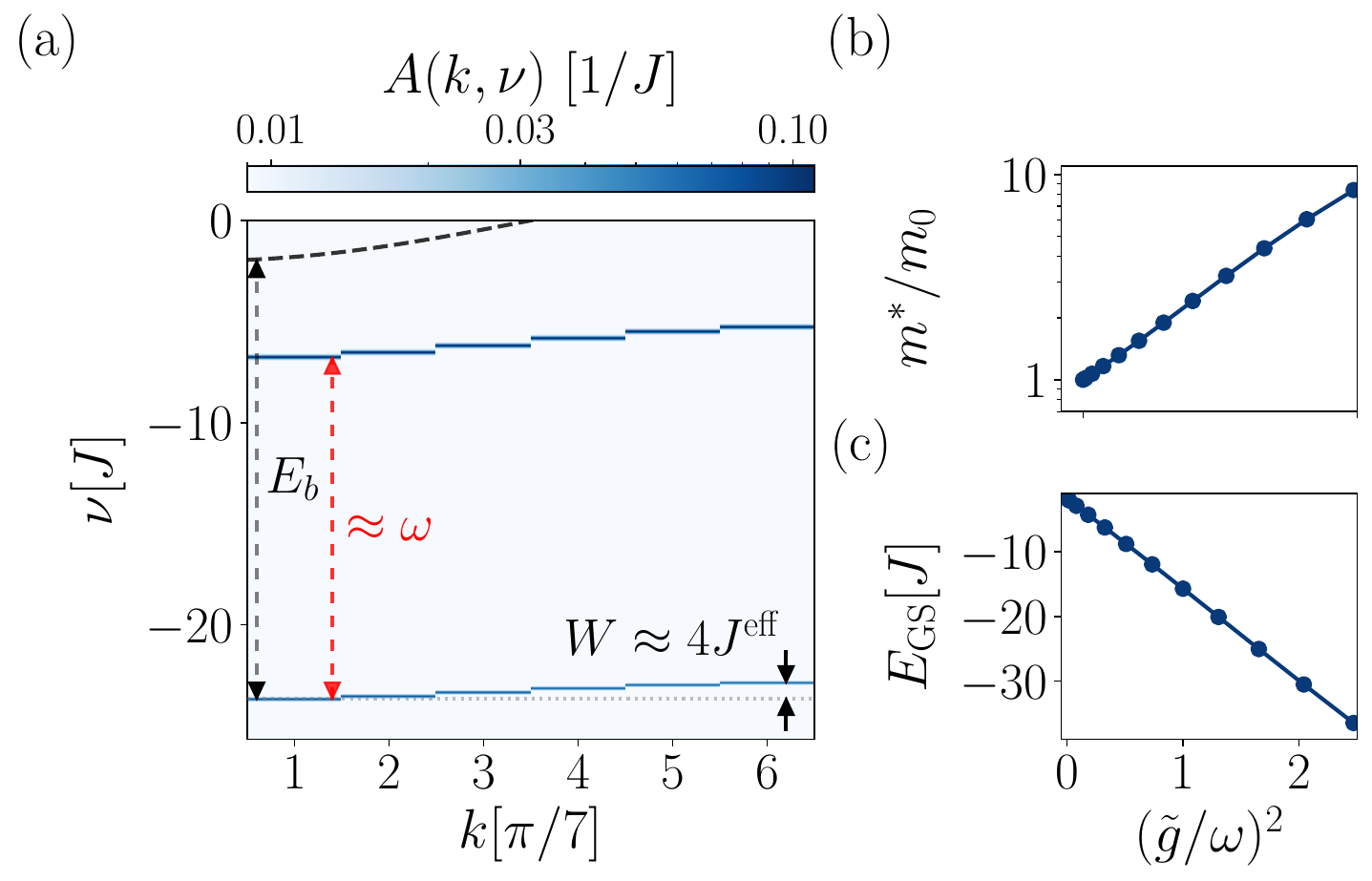}
    \caption{ 
    (a) Spectral function $A(k,\nu)$ of the on-site Holstein polaron for $\tilde{g}/\omega = 1.25$ and $\omega/J = 15$. We compute the spectral function using Eq.~\eqref{eq:spectral_function} and replacing the Dirac delta function with a Lorentzian with width $\gamma/J = 0.05$. $k$ represents the standing-wave quasi-momentum (see Appendix \ref{sec:app-obc}). The color bar indicates the amplitude of the spectral function on a log scale for different frequencies and momenta. The dashed black line indicates the free-spin dispersion. The dashed black arrow represents the polaron binding energy, $E_b = E_{\rm GS} - E_{\rm GS}\,(\tilde{g} = 0)$, and the dashed red arrow indicates the separation of the single-phonon excitation, $\omega$. The two solid black arrows on the right indicate the bandwidth of the system, which corresponds to approximately $4J^{\rm eff}$. 
    (b) Renormalization of the effective mass, $m^*/m_0$, of the Holstein polaron as a function of coupling $(\tilde{g}/\omega)^2$ and $\omega/J = 15$ (see Sec.~\ref{sec:LF}).
    (c) Polaron ground-state energy $E_{GS}/J$ as a function of coupling $(\tilde{g}/\omega)^2$ and $\omega/J = 15$.}
    \label{fig:Holstein-spectral}
\end{figure}

{\bf\emph{(i)}} \emph{Polaron cloud and binding energy:} Fig.~\ref{fig:Holstein-spectral} shows the spectral function for the nearest-neighbor on-site Holstein model, revealing several key signatures of polaron formation. The sharp, well-defined individual peaks at discrete momentum values confirm that $k$ remains a good quantum number despite strong electron-phonon coupling. This indicates that the polaron retains its coherent, dispersive character as a well-defined quasiparticle. 
The polaron ground-state energy $E_p$ is shifted below the bare particle energy by a binding energy $E_b \sim -g^2/\omega$, reflecting the attractive self-induced potential from the surrounding phonon cloud (see Fig.~\ref{fig:Holstein-spectral}(a)).
Using the fact that phonons are in the vacuum state within the Lang-Firsov frame, we can approximate the ground-state wavefunction as
\begin{equation}
    \begin{split}
        \ket{\psi_{\rm GS}} &\approx U^\dag_{\rm LF} \sum_i c_i \sigma_i^+ \ket{\downarrow}_N \otimes \ket{0_{\rm ph}} \\
        &=  \sum_i c_i \sigma_i^+ \ket{\downarrow}_N \otimes \ket{\alpha_i^{LF}},
    \end{split}
    \label{eq:polaron_gs}
\end{equation}
where $c_i$ are complex coefficients, and $\ket{\alpha_i} =  \exp\left(\sum_\lambda \alpha^{LF}_\lambda(i) a_\lambda^\dag - {\alpha^{LF}_\lambda}^*(i) a_\lambda\right) \ket{0}$ is a coherent state with displacement $\alpha^{LF}_\lambda(i) = \tilde{g}_{i\lambda}/\omega_\lambda)$, representing the phonon cloud localized around the spin excitation at site $i$. Evaluating the energy of the effective Hamiltonian in Eq.~\eqref{eq:eff_hamiltonian} shows that the polaron ground-state energy at strong coupling scales as $E_p \propto -\tilde{g}^2/\omega$, with corrections of order $J$ (see Fig.~\ref{fig:Holstein-spectral}(c)), reflecting the polaron binding energy between spins and phonons. 

{\bf\emph{(ii)}} \emph{Effective mass:} The lowest-energy band corresponds to the polaron ground state, whose dispersion allows us to extract the effective mass $m^*$ (see Fig.~\ref{fig:Holstein-spectral}(b)). As shown in Eq.~\eqref{eq:eff_hopping}, the effective hopping gets exponentially suppressed with increasing coupling. Since the NN hopping gives $E(k) = 2 J^{\rm eff} \cos(k)$, the exponential suppression of $J^{\rm eff}$ leads to an exponential increase of the effective mass, $m^* = 1/ \frac{\partial^2 E}{\partial k^2}|_{k = 0} \sim 1/J^{\rm eff}$ and an exponential narrowing of the bandwidth, $W \approx 4J^{\rm eff}$. In addition, the magnitude of the hopping renormalization depends on the exact form of spin-phonon coupling. For optical phonon modes considered here,  the coupling can be rewritten in real space $\tilde g_{im} = \tilde {g}_{i-m}$. Therefore, the renormalization depends on the difference of coupling between the neighboring sites $i,j = i\pm 1$
 \begin{equation}
    J_{ij}^{\rm eff} = J_{ij} \exp\left( - \frac{1}{2} \sum_m \left|\frac{\tilde {g}_{i-m} - \tilde {g}_{j-m}}{\omega}\right|^2\right).
    \label{eq:eff_hopping_local}
\end{equation}
For spin-phonon couplings $\tilde{g}_{i-m}$, which decay faster with distance $|i-m|$, the hopping renormalization gets stronger, and the effective mass becomes larger. 
The renormalization of the hopping can be understood by the phonon dressing of the particle. In the Holstein model, the electron–phonon coupling is short-ranged, so the polaron cloud is strongly localized on a single site. When the particle hops to a neighboring site, the lattice distortion must rearrange, and the hopping amplitude is reduced by the overlap of the initial and final phonon configurations, the so-called Franck–Condon factor defined by the exponential factor in Eq.~\eqref{eq:eff_hopping_local}. This overlap is exponentially small in the coupling strength, leading to a strong exponential suppression of the hopping $\sim 1/m^*$. In Sec.~\ref{sec:dynamical_mass_renorm}, we describe a protocol for extracting the effective mass of the polaron by tracking single particle dynamics of a spin-phonon polaron Hamiltonian in the Lang-Firsov regime. 

{\bf\emph{(iii)}} \emph{Single-particle residue:} The change in the spectral weight of the lowest band $|\langle\psi_{\rm polaron}|j\rangle|^2$, which measures the overlap between the dressed polaron and bare spin excitation, signals the strong spin-phonon entanglement that characterizes the polaronic ground state. To quantify how strongly the spin excitations get dressed by the polaron background, we define the local single-particle residue 
\begin{equation}
    Z_j =  |\braket{\psi_{\rm GS}|j}|^2. 
\end{equation}
Using the expression for the ground state wavefunction in Eq.~\eqref{eq:polaron_gs}, we see that the residue gets suppressed by the overlap between the phonon cloud and the phonon vacuum. In the case of Lang-Firsov approximation, this corresponds to a coherent state, resulting in exponential suppression with the amplitude of the coherent state and the coupling strength squared, $Z_j \sim e^{-\tilde{g}^2/\omega^2}$.
We explore the quasiparticle weight renormalization in greater details in Sec.~\ref{sec:displaced_ldos}, where we show that individual matrix elements of the local density of states exactly correspond to the single-particle residue. 

{\bf\emph{(iv)}} \emph{Single-phonon excitations:} Above the main polaron band, we observe a one-phonon excited state at an energy approximately $\omega$ above the polaron ground state. 
While the effective Hamiltonian in Eq.~\eqref{eq:eff_hamiltonian} was derived assuming the phonon vacuum, the Lang–Firsov transformation also allows us to systematically analyze low excited states of the system (see Appendix \ref{sec:app-Lang-firsov} for details). In particular, the first excited band of the Holstein model is a single-phonon excitation on top of the ground state polaron, with energy approximately $E_p + \omega$. The phonon sideband provides direct spectroscopic evidence of the composite spin-phonon nature of the polaron ground state, since the evolution of the spin generates phonon sidepeaks. In Sec.~\ref{sec:displaced_ldos}, we show how the phonon sidebands can be observed for a general spin-phonon polaron Hamiltonian.

\section{Long-range hopping polaron: Displaced local density of states and Polaron tomography}
\label{sec:LDOS+Tomography}

We begin by establishing a set of observables for characterizing polaron physics in trapped-ion quantum simulators. We circumvent the problem of measuring the ground state through two complementary approaches. First, we identify observables, including spectral functions, local density of states, and correlation functions, which can be extracted from real-time dynamics via hybrid-digital protocols using ancilla ions initially entangled with the system. Second, we demonstrate that preparing the system in displaced product states motivated by the Lang-Firsov transformation significantly improves overlap with the ground state, enabling direct observation of equilibrium polaron signatures. Generalizing the preparation to include arbitrary coherent states or displaced Fock states (see Sec.~\ref{sec_dfock_GF}) further enables \textit{polaron tomography}.

\subsection{Local density of states}

While the spectral function provides complete momentum-resolved information, it requires measuring the Green's function between all pairs of sites and is only meaningful when momentum is a good quantum number. For finite systems with open boundaries and long-range hopping or spin--phonon coupling, the standing-wave quasi-momentum generally no longer provides a convenient quantum number for the full interacting problem. It is therefore more convenient to consider the local density of states (LDOS) given by
\begin{align}
    \rho(\nu) &= - \frac{1}{\pi }\sum_j \Im{G(j,j,\nu)}\nonumber\\ &= \sum_j \sum_n|\langle \psi_n \ket{j}|^2  \delta(\nu - E_n),
    \label{eq:ldos}
\end{align}
where $G(j,j,\nu)$ is a Fourier transform of $G(j,j,t)$. Similar to the spectral function, the LDOS still captures the system's full energy spectrum, but is more experimentally accessible. 
\begin{figure}
    \centering
    \includegraphics[width=1.0\linewidth]{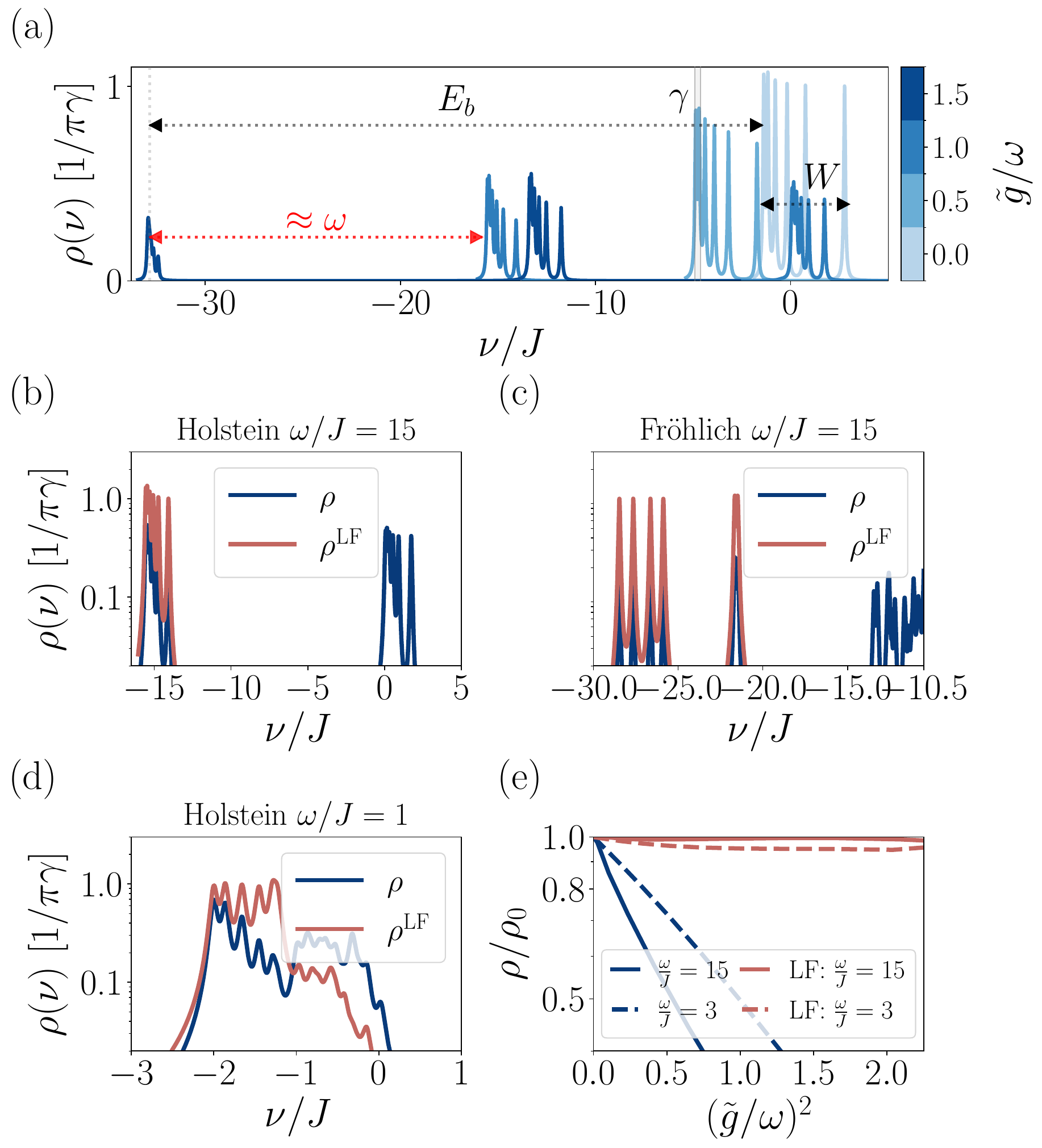}
    \caption{
    {\bf Bare and displaced local density of states.} 
    (a) Local density of states of the on-site Holstein model with long-range hopping for different values of coupling strengths $\tilde{g}/\omega$, with $\alpha_J = 1.2$ and $\omega/J = 15$. The dashed black arrow represents the polaron binding energy, $E_b = E_{\rm GS} - E_{\rm GS}\,(\tilde{g} = 0)$, and the dashed red arrow indicates the energy gap to the first phonon sideband, which is $\approx\omega$. $W$ indicates the bandwidth of the system. The vertical gray lines indicate the linewidth $\gamma/J = 0.05$.
    (b)--(d) Comparison of the bare LDOS (blue), Eq.~\eqref{eq:ldos}, and displaced LDOS (red), Eq.~\eqref{eq:ldos_disp}, for long-range hopping with $\alpha_J = 1.2$, $\tilde{g}/\omega = 1$, and linewidth $\gamma/J = 0.05$. (b) On-site Holstein model with $\omega/J = 15$. 
    (c) Fr\"ohlich model with $\alpha = 1.6$ and $\omega/J = 15$. (d) On-site Holstein model with $\omega/J = 1$. 
    (e) Comparison of the ground-state peak height between the bare LDOS (blue) and displaced LDOS (red) for on-site Holstein model with $\omega/J = 15$ (solid) and $\omega/J = 3$ (dashed) with respect to the coupling strength $(\tilde{g}/\omega)^2$. $\rho_0$ is the density of states of the lowest peak at zero coupling, and $\gamma/J = 0.05$.}
    \label{fig_ldos_disp_Hol}
\end{figure}
The LDOS for the on-site Holstein model with long-range hopping at various coupling strengths is shown in Fig.~\ref{fig_ldos_disp_Hol}(a). At zero coupling (blue line), the spectrum consists of multiple peaks, forming a wide band, centered around zero frequency. Since individual peaks correspond to eigenenergies of the system, the total spread of the peaks arises from the dispersion of the ground states and corresponds to the bandwidth of the system $W$. The individual peak width is given by phenomenological broadening $\gamma$, which depends on the total evolution time
for the Green's function measurement (see Sec.~\ref{sec:protocols} for details). 

As the coupling increases, the lowest polaron band exhibits four main features (See Fig.~\ref{fig_ldos_disp_Hol}): a shift to more negative energies, a decrease in width, a decrease in height, and the appearance of sidebands. The shift of the peaks to lower energies corresponds to phononic dressing of the spin and shifts by the polaron binding energy $E_b \sim -\tilde{g}^2/\omega$ (see Fig \ref{fig:Holstein-spectral}(c)). 
In the case of finite coupling and $\omega > 2J$, the width of the lowest band results from the interplay of the renormalized hopping $J^{\rm eff}$
and the effects of the static local potential generated by the phonon cloud (second term in Eq. \eqref{eq:eff_hamiltonian}). Particularly, the exponential renormalization of the hopping $J^{\rm eff}$ reduces the overall bandwidth $W$ of the system, while the static potential results in the splitting of the band. In the following section, we discuss protocols for extracting effective hopping directly in the case of strongly coupled systems. 

The decrease in height of the lowest peak arises from the quasi-particle renormalization. Given a finite broadening $\gamma$, the Dirac delta function $\delta(\nu - E_n)$ in Eq.~\eqref{eq:ldos} becomes a Lorentzian with broadening $\gamma$, with the amplitude proportional to the single-particle residue
\begin{equation}
    \rho(E_{\rm GS})  = \frac{1}{\gamma \pi} \sum_j |\bra{\psi_{GS}} j \rangle|^2 = \frac{1}{\gamma \pi} \sum_j Z_j.
\end{equation}
As the coupling is increased, the single-particle residues decrease exponentially, reflecting the dressing of the spin by the phonon cloud and resulting in a decrease of peak height. We note that the single-particle residue can also be defined for arbitrary eigenstates by considering $\rho(E_n)$. Since the whole ground-state band corresponds to phonon vacuum within the LF frame, the weight of the whole band decreases exponentially with the coupling strength. 
As the coupling increases, the spacing between individual levels can fall below the broadening $\gamma$, thereby reducing the resolvability of individual peaks (see Fig.~\ref{fig_ldos_disp_Hol}(a) for $\tilde{g}/\omega = 1.5$). In this case, integrating the LDOS over the first band gives the band-averaged single particle residue $ \int_{E_{\rm GS}}^{E_{\rm GS} + W} d\nu \rho(\nu)  = \sum_{n \in \rm{Band}} \sum_j Z^{n}_j = N_{\rm s}   Z^{\rm Band}$, which still shows exponential renormalization.  Finally, at strong coupling, sidebands appear, corresponding to single-phonon excitations at energy $\omega$ above the ground state. 

Overall, the main qualitative features characterizing the polaron formation can be extracted from the local density of states, without having to measure the full Green's function between all pairs of points. Furthermore, our results also demonstrate that the main polaronic features present in the case of the NN hopping Holstein model also survive in the long-range hopping, thereby preserving key polaronic signatures (blue lines in Fig.~\ref{fig_ldos_disp_Hol}(b)--(d)). 

\subsection{Displaced local density  of states} \label{sec:displaced_ldos}

We now introduce \textit{displaced local density of states}, a generalization of the local density of states that provides further information about the structure of the polaron wavefunction. The key idea is to prepare each phonon of the system in a coherent state, instead of the vacuum state (for details, see Sec.~\ref{sec:protocols}). The specific form of each coherent state can be tuned to match the expected phonon cloud surrounding the spin excitation, allowing us to directly probe the composite spin-phonon structure of the polaron. We define the displaced real-time Green's function as
\begin{equation}
    G^{\alpha\beta}(j,l, t) = -i \theta(t) \bra{l} D(\beta) e^{-iHt} D(\alpha) \ket{j},
\label{eq:displacedGF}
\end{equation}
where the initial state and the final states are each an outer product of a localized spin excitation and a combination of coherent states of phonons created by the displacement operator
\begin{equation}
    D(\alpha) = \exp\left(\sum_m \alpha_m(j) a_m^\dag - \alpha_m^*(j) a_m\right),
\end{equation}
that can be generated using ancilla ions~\cite{Whitlow2023}. In the following, we choose the displacement $\alpha_\lambda(j)$ based on the position at which the spin is initially created. The phonons must be displaced differently depending on the spin position $j$ in the initial product state, as this allows the construction of the expected phonon cloud for a spin at site $j$. Crucially, this protocol does not require entangling phonon and spin degrees of freedom at the initial time.  Following the same procedure as for the standard LDOS and setting $\beta_m(j) = -\alpha_m(j)$, the Fourier transform of the on-site displaced Green's functions $G^{\alpha}(j,j,t)$ results in the displaced local density of states (dLDOS)
\begin{equation}
    \begin{split}
            \rho^\alpha(\nu) &= -\frac{1}{\pi}\sum_j \Im{ G^{\alpha}(j,j,\nu)}\\ 
            &= \sum_{j,n} |\langle \psi_n | D(\alpha(j))| j\rangle|^2  \delta(\nu - E_n).
    \end{split}
    \label{eq:ldos_disp}
\end{equation}
Since we know the approximate form of the wavefunction in the strong-coupling limit \eqref{eq:polaron_gs}, we choose the displacement appropriately as
\begin{equation}
    \alpha^{\rm LF}_m(j) = \frac{\tilde{g}_{jm}}{\omega_m},
    \label{eq:LF-displacement}
\end{equation}
which corresponds to the Lang-Firsov displacement. In Fig. \ref{fig_ldos_disp_Hol}(b)--(d), we show the dLDOS using a single coherent state $\alpha^{\rm LF}$. The $n$-th eigenstate of the system can generally be written as $\sum_j c_j^n \sigma_j^+ \ket{\downarrow}_N \ket{\phi_j^n}$, with $c_j^n$ being complex coefficients and $\ket{\phi_j^n}$ corresponding to a general phonon wavefunction. If the initial displaced state corresponds to the exact ground-state cloud (see Eq.~\eqref{eq:ldos_disp}),  the overlap $|\langle \phi_j^{\rm GS} | D(\alpha(j)) | 0_{\rm ph} \rangle|^2$ with the ground-state phonon wavefunction approaches unity, due to the renormalization of $c_i^n$ coefficients. Moreover, from the Lang-Firsov analysis, we expect the excited states to consist primarily of few-phonon excitations on top of the Lang-Firsov displaced state, suggesting that the Lang-Firsov displacement provides the natural starting point for understanding the structure of polaronic excitations.

Fig.~\ref{fig_ldos_disp_Hol}(b)--(c) compares the bare and displaced LDOS for the Holstein and Fr\"ohlich models. The support of both spectral functions is identical, since they both reflect the same energy eigenstates. The important difference lies in the spectral weights. The undisplaced LDOS exhibits much smaller spectral weight for the lowest band, reflecting the vanishing quasi-particle weight $|\langle\psi_{\rm GS}|i,0_{\rm ph}\rangle|^2$ between bare and dressed polaron states. By contrast, the Lang-Firsov displaced LDOS maintains high spectral weight across all coupling strengths, confirming that $\ket{i,\alpha^{\rm LF}_i}$ closely captures the ground-state phonon cloud structure. Looking at the excited spectrum, we see that the phonon sidebands vanish in the Lang-Firsov dLDOS, showing that the initial state is orthogonal to the first excited band. This implies that the initial state closely captures the ground state, since the majority of the phonon cloud is correctly captured. Therefore, the displaced density of states can be used to characterize the phonon clouds corresponding to individual polaronic bands. 

Fig.~\ref{fig_ldos_disp_Hol}(d) compares the LDOS and dLDOS for the on-site Holstein model in the intermediate regime with $\omega/J = 1$. In this regime, the Lang-Firsov picture is no longer straightforwardly applicable, since the spin and phonon energy scales are comparable. Nevertheless, the dLDOS still shows higher spectral weights, demonstrating that the Lang-Firsov frame provides a natural starting point for understanding the intermediate regime by connecting it to both the weak-coupling and ultrastrong-coupling limits (see Fig.~\ref{fig_regimes}). Moreover, describing the spectral weights as a function of the displaced state enables the characterization of the phonon content of the polaron cloud. In the next subsection, we take this idea further and describe a protocol to extract the full polaron wavefunction, which we refer to as \textit{polaron tomography}.

Fig.~\ref{fig_ldos_disp_Hol}(e) shows the spectral weight of the LDOS and dLDOS as a function of coupling strength for $\omega/J =15$ and $\omega/J = 3.0$. In the anti-adiabatic case with $\omega/J = 15$, the weight of LDOS decreases exponentially with coupling strength, reflecting the phonon cloud dressing of the single spin excitations and the decay of single particle residue $Z_i$. On the contrary, the dLDOS with Lang-Firsov displacement displays constant weight, proving that the Lang-Firsov wavefunction describes the ground state of the system well. For the $\omega/J = 3.0$ case, both spectral functions initially decay slowly with increasing coupling. However, at stronger coupling, the bare LDOS undergoes exponential suppression, while the dLDOS recovers.
This can be understood as follows. At small coupling, the weight renormalization is perturbative, resulting in small renormalization of the weight. When the coupling is increased, and the system reaches an intermediate regime $\tilde{g} \sim \omega \sim J$, the ground state significantly differs from the Lang-Firsov state, and thus the weight is reduced. As the coupling is further increased, however, the system approaches the regime where $\omega \gg J^{\rm eff}$, and the Lang-Firsov approximation becomes applicable again. The Lang-Firsov dLDOS recovers the weight back to unity, while the bare LDOS continues decaying exponentially, as in the $\omega/J = 15$ case. Therefore, the Lang-Firsov dLDOS can help quantify the values of parameters at which the system deviates the most from the Lang-Firsov approximation and the pertubative regime, allowing for identification of the intermediate regime.  The dLDOS also has important practical implications, as it provides coupling-independent spectral weight, which allows the extraction of spectral signatures without the exponential decay of the signal strength with increasing coupling strength. We note that the extraction of displaced Green's function is unique to trapped-ion systems and goes beyond traditional condensed matter probes.

\subsection{Polaron Tomography}

We now highlight another feature of our displaced technique, called \textit{polaron tomography}. The full displaced Green's function can be written using the spectral decomposition
\begin{equation}
    \begin{split}
        G^{\alpha\beta}(j,l,\nu) &= \sum_n \frac{\bra{l,\beta_j} \psi_n \rangle \langle \psi_n \ket{j,\alpha_i}}{\nu - E_n + i0^+},  \\
    \end{split}
\end{equation}
where $n$ indicates the $n$-th eigenstate of the system. The displaced Green's function provides direct access to the density matrix of individual polaron eigenstates in a chosen basis. For an eigenstate $\ket{\psi_n}$, the density matrix is defined as $\rho_n = \ket{\psi_n}\bra{\psi_n}$. In the basis of localized spin excitations accompanied by displaced phonon
states, the matrix elements of this density matrix are
\begin{equation}
    \rho^{(n)}_{(l,\beta),(j,\alpha)}
    =
    \bra{l,\beta_j}\psi_n\rangle
    \langle\psi_n\ket{j,\alpha_i}.
\end{equation}
These matrix elements appear directly in the spectral decomposition of the
displaced Green's function,
$
    G^{\alpha\beta}(j,l,\nu)
    =
    \sum_n
    \rho^{(n)}_{(l,\beta),(j,\alpha)}
    /
    {(\nu-E_n+i0^+)}
$.
By tuning the probe frequency close to an isolated resonance,
$\nu \approx E_n$, the contribution from the corresponding eigenstate
dominates the Green's function, while all other terms are suppressed by their
finite energy denominators. Therefore, extracting the residue of the Green's function at the resonance
directly yields the individual matrix elements of the density matrix of the
polaron eigenstate,
\begin{equation}
    \rho^{(n)}_{(l,\beta),(j,\alpha)}
    =
    \lim_{\nu\rightarrow E_n}
    (\nu-E_n+i0^+)
    G^{\alpha\beta}(j,l,\nu).
\end{equation}
Repeating this procedure for different positions of spin excitations and coherent states displacements reconstructs the density matrix in the displaced basis. Since coherent states form an overcomplete representation of the phonon Hilbert space, this provides a redundant but complete characterization of the spin--phonon wavefunction. For a non-degenerate eigenstate, the reconstructed density matrix is a rank-one projector, and its unique non-zero eigenvector gives the expansion coefficients of the polaron wavefunction in the chosen basis, up to an overall phase. 

Here, we primarily discussed using the coherent states as a basis. However, the phonon basis of choice can also be optimized to more efficiently extract the polaron matrix elements. For example, the displaced basis built on the Lang-Firsov transformation (see Eq. \eqref{eq:LF-displacement}) could be accompanied by $n$ phonon excitation. This displaced Fock basis could be realized in ions with sequences of spin-dependent forces and resonant sideband excitations~\cite{leibfried2003quantum, pagano2025varenna}. Since deviations from the Lang-Firsov approximation can be most easily quantified in terms of a few-phonon excitation on top of the polaron state,  this displaced Fock basis could be a natural basis for enhancing the overlap with excited states, providing a promising pathway for probing the full phonon wavefunction in the intermediate regime (see Sec. \ref{sec_dfock_GF}). 

\section{Dynamical extraction of mass renormalization}
\label{sec:dynamical_mass_renorm}

Having discussed the spectral properties of the polaron system, we now turn to dynamical expectation values, focusing on spin-occupation dynamics. To demonstrate the protocol in the simplest scenario, we consider the NN hopping Holstein and Fr\"ohlich models. The realization of approximately NN hopping is possible in trapped-ion systems by detuning the laser drive used to generate spin-spin interactions close to the lowest radial mode of motion and using alternating optical phases on neighboring ions \cite{de2024observationstringbreakingdynamicsquantum, luo2025quantumsimulationbubblenucleation}.

\begin{figure}[h]
    \centering
    \includegraphics[width=1.0\linewidth]{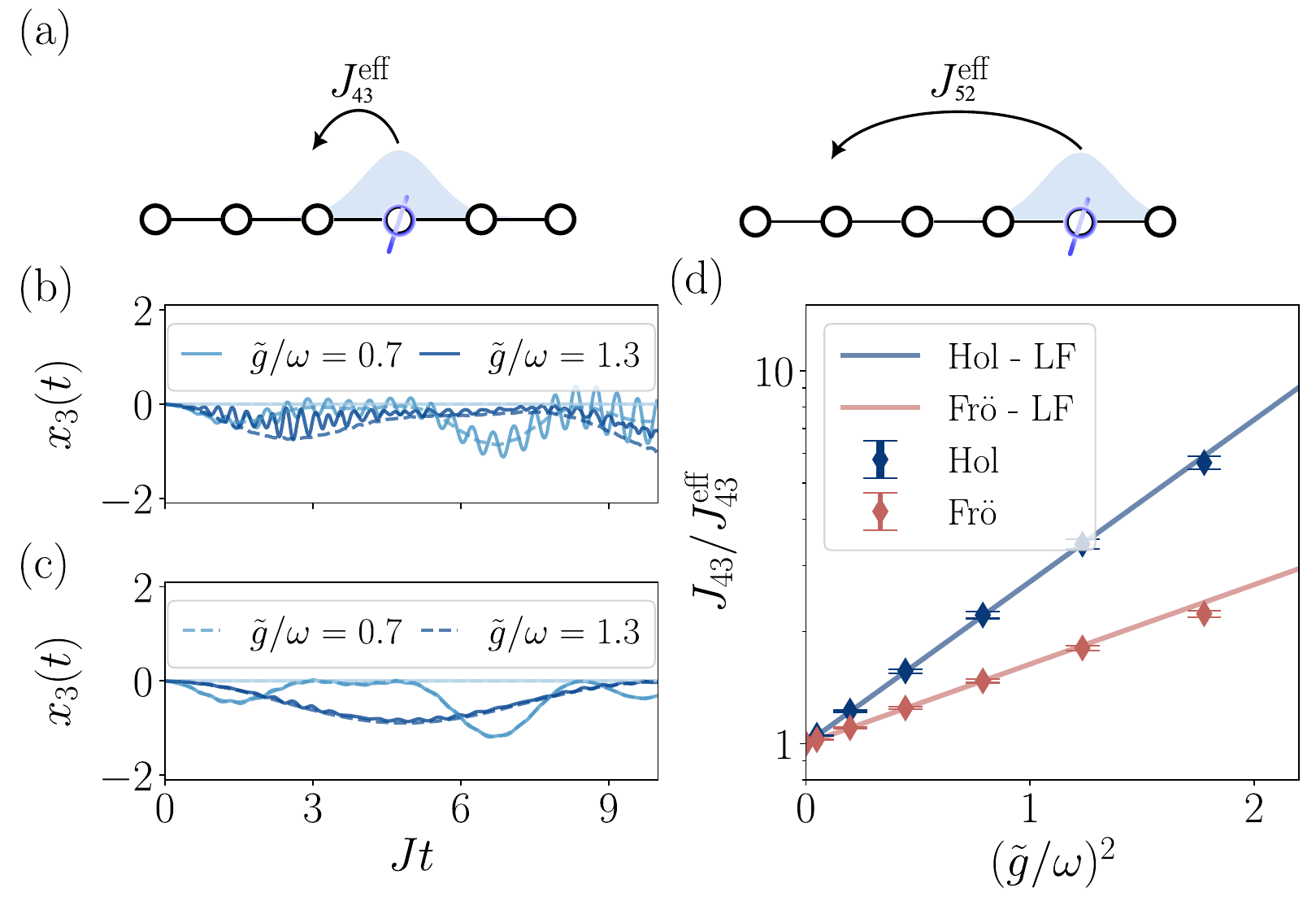}
    \caption{(a) Schematic of phonon-dressed effective spin hopping. A spin excitation (blue) is accompanied by a local phonon displacement (purple envelope), while the arrows illustrate effective hopping matrix elements $J^{\rm eff}_{ij}$ between different ion sites, with $J^{\rm eff}_{43}$ and $J^{\rm eff}_{52}$ shown as examples. (b)--(c) Real-time evolution of phonon displacement $ x_{m} = \langle a_m + a_m^\dag\rangle$ at site $m = 3$. The spin-phonon coupling Hamiltonian corresponds to an on-site Holstein model with nearest-neighbor hopping and $\omega/J = 15$ for different values of $\tilde{g}/\omega$. The dashed lines indicate the adiabatic displacement relative to the spin $2 \tilde{g}/\omega \cdot  \expval{n_3}(t)$. The initial state is prepared in site $4$ with (b) phonon vacuum and (c) displaced phonon background $\alpha^{\rm LF}$. (d) Extracted effective hopping versus coupling strength $(\tilde{g}/\omega)^2$ with $\omega/J = 15$ for NN hopping, on-site Holstein model (blue), and Fr\"ohlich model with $\alpha = 3.0$ (red). The lines indicate the analytic Lang-Firsov prediction for the renormalization of the $J_{43}$ hopping element.}
    \label{fig:effective_hopping_nn}
\end{figure}

The dynamics of a single spin involve multiple timescales set by the competing terms in the Hamiltonian, making it difficult to extract individual ground-state polaron properties, such as the effective hopping $J^{\rm eff}$, from the real-time evolution. This can be directly seen in Fig.~\ref{fig:effective_hopping_nn}(b). We initialize a spin excitation at the right central site (site $4$) with phonons in vacuum and track the occupation, as well as the phonon displacement at the neighboring left central site (site $3$). As the coupling increases, the phonon displacement exhibits rapid oscillations superimposed on a slower envelope, while the spin evolution slows down. The interplay between the fast phonon oscillations and slow spin transport makes it difficult to cleanly extract the individual timescales and thus the ground-state polaron properties. 

To isolate the slow spin dynamics, we instead initialize the system in a displaced phonon state: a spin localized at site 4 surrounded by a Lang-Firsov displaced phonon cloud, $D(\alpha^{\rm LF})\ket{4,0_{\rm ph}}$, as defined in Eq.~\eqref{eq:LF-displacement} (see Fig.~\ref{fig:effective_hopping_nn}(a)). Fig.~\ref{fig:effective_hopping_nn}(c) demonstrates the dramatic effect of displacing the initial state. The rapid phonon oscillations vanish, and the phonon displacement smoothly tracks the spin position. Crucially, the spin dynamics at different couplings have identical functional form, differing only in their characteristic timescale, indicating that the dynamics depends only on a single energy scale, corresponding to the effective spin hopping $J^{\rm eff}$.  

The success of the protocol can be understood by considering the energy scales in the problem and the nature of the initial state. The initial displacement ensures that, \textit{within the Lang-Firsov frame}, the initial state corresponds to the phonon vacuum, $U_{\rm LF} D(\alpha^{\rm LF}) \ket{4,0_{\rm ph}} = \ket{4}_{\rm LF}$. Since the lowest polaron band corresponds to phonon vacuum in the LF frame (see Sec.~\ref{sec:LF}), the initial state corresponds to the lowest polaron band.  
In the $J^{\rm eff} \ll \omega$ limit, the system cannot generate additional phonons within the Lang-Firsov frame. Therefore, the system starts and remains in a state corresponding to the lowest polaron band. As a result, the real-time evolution is dominated by the lowest band dynamics, which is described by a single energy scale $J^{\rm eff}$. This means that renormalization of $J^{\rm eff}$ simply renormalizes the timescale of the dynamics, which can be extracted to obtain $J^{\rm eff}$.

\subsection{Extracting the effective mass}

By fitting the effective time scale of the spin-occupation curves, we extract the effective hopping timescale $J^{\rm eff}(g)$ as a function of coupling strength. Fig.~\ref{fig:effective_hopping_nn}(d) shows the extracted hopping renormalization for both Holstein and Fr\"ohlich models. The results are in agreement with the Lang-Firsov analytical prediction, demonstrating the validity of our protocol. Comparing the two models, we observe that the Holstein effective hopping gets renormalized more strongly than the Fr\"ohlich one, consistent with the more localized nature of Holstein coupling. Both models show exponential renormalization of the hopping strength, as expected from the Lang-Firsov picture. 

We note that attempting to extract $J^{\rm eff}$ from bare initial state dynamics without the displacement $\ket{4,0_{\rm ph}}$ would yield inaccurate results. Without the displaced initial state, the dynamics are much more involved, mixing ground-state and excited-state contributions, making it difficult to extract the effective hopping timescale. The protocol implemented above relies on the spin evolution between sites $4$ and $3$, allowing the extraction of the effective hopping $J_{43}^{\rm eff}$. We choose these sites to minimize the edge effects arising from finite system size and to remove any effects which might arise from the static background potential, particularly relevant for the Fr\"ohlich model (see Sec.~\ref{sec:LF}), where the phonons generate an effective local potential. Since the systems we consider, as well as experimental trapped-ion realizations, possess inversion symmetry, the static potential does not affect the initial dynamics between two middle sites (sites $4$ and $3$).

\section{Experimental Scheme for polaron dynamics}

\label{sec:Long-range_hopping_polaron}

In the following section, we discuss how the key polaron signatures can be observed on trapped-ion systems with spin--phonon couplings optimized to engineer Holstein or Fr\"ohlich models (see Sec.~\ref{sec:Engineering_electron-phonon_models}). We focus on the case of long-range hopping with $\alpha_J = 1.2$ and optical phonons with $\omega/J = 5$.  We consider the `ideal' case of the Holstein model with 6 spin sites, coupled to 6 local phonon modes, and the `trapped-ion' case of 6 spin sites coupled to 8 normal phonon modes, with couplings optimized to mimic the 6 local phonon modes of the ideal model using the RLP mapping. To compare the two models, we introduce the dimensionless coupling
\begin{equation}
    \Lambda = \frac{1}{N_{\rm s}} \sum_{i,m} \frac{\tilde{g}^2_{im}}{\omega_m J},
\end{equation}
where $N_{\rm s} = 6$ is the number of spin sites. In the case of `ideal' models, the index $m$ corresponds to the local mode index $m$ (defined by a map described in Sec.~\ref{sec:Engineering_electron-phonon_models}), and the parameter $\Lambda \sim \tilde{g}^2/J\omega$ approximately quantifies the polaron binding energy $E_b$, as discussed before. 

\subsection{Local density of states}

\begin{figure}[h]
    \centering
    \includegraphics[width=1.0\linewidth]{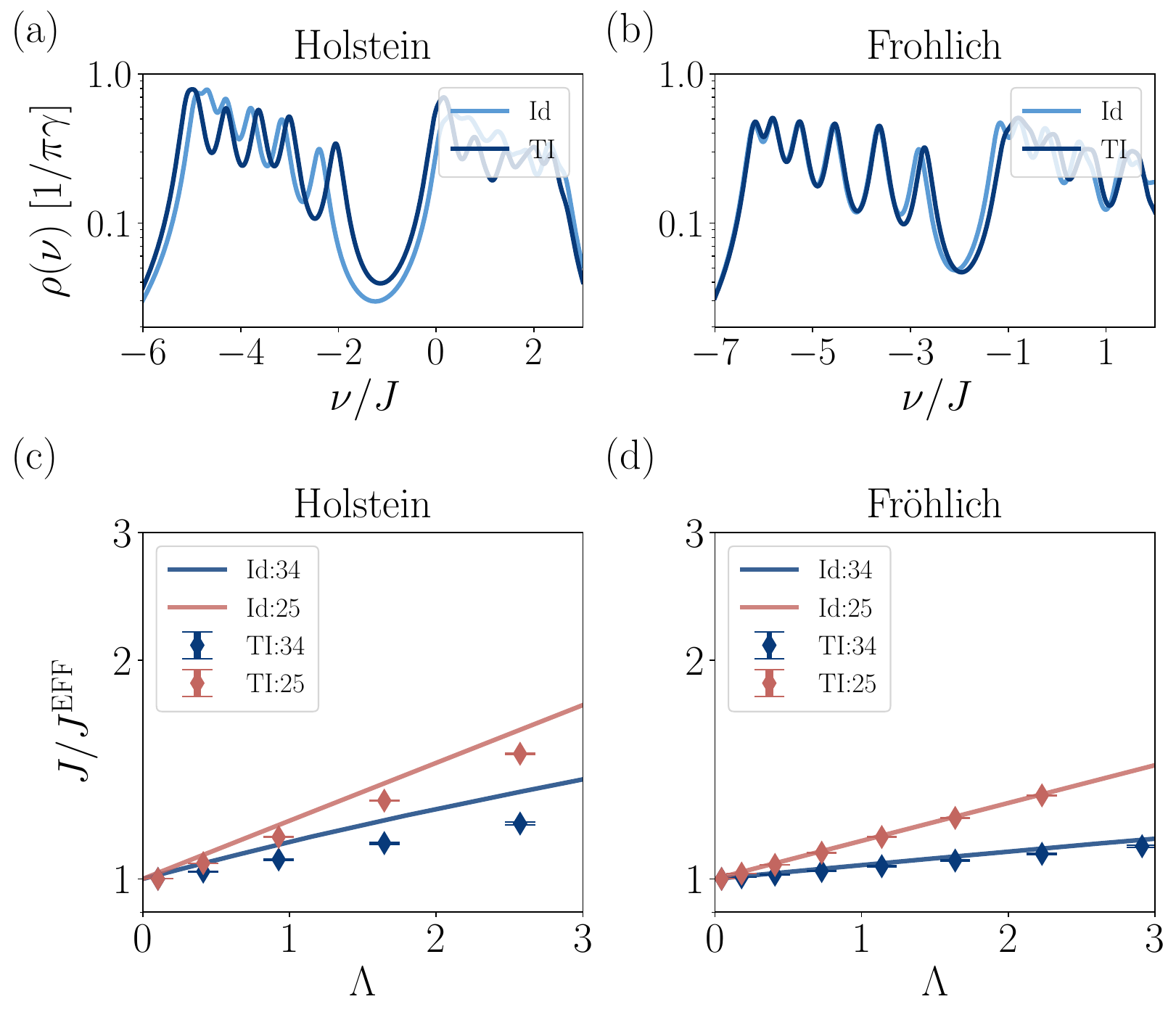}
    \caption{
    Comparison of trapped-ion and ideal spin--phonon models. (a) Local density of states for long-range hopping Holstein model with $\xi = 0.83$, $\omega/J = 5$, $\Lambda = 3.7$, and $\alpha_J = 1.2$ for ideal (light blue) and trapped-ion coupling constants (dark blue). (b) Local density of states for long-range hopping Fr\"ohlich model with $\alpha = 1.6$, $\omega/J = 5$, $\Lambda = 4.6$, and $\alpha_J = 1.2$ for ideal (light blue) and trapped-ion coupling constants (dark blue). In both (a) and (b), the bosonic cutoff is $4$ per mode, and the linewidth of $\gamma/J = 1/6$ was used. (c) Effective hopping renormalization of $J_{43}$ (blue) and $J_{52}$ (red) for the long-range hopping Holstein model with $\xi = 0.83$, $\omega/J = 5$, and $\alpha_J = 1.2$ as a function of coupling $\Lambda$ for ideal (full line) and experimental coupling constants (diamonds). (d) Effective hopping renormalization of $J_{43}$ (blue) and $J_{52}$ (red) for the long-range hopping Fr\"ohlich model with $\alpha = 1.6$, $\omega/J = 5$, and $\alpha_J = 1.2$ as a function of coupling $\Lambda$ for ideal (full line) and experimental coupling constants (diamonds). In (c) and (d), the bosonic cutoff is $5$ per mode.}
    \label{fig:long-range_hol_vs_fro}
\end{figure}

We show the LDOS for the ideal and trapped-ion Holstein model for $\xi = 0.83$ and Fr\"ohlich model for $\alpha = 1.6$ and long-range hopping with $\alpha_J = 1.2$ in Figs.~\ref{fig:long-range_hol_vs_fro}(a) and \ref{fig:long-range_hol_vs_fro}(b), respectively. In both cases, a spectral broadening of $\gamma/J = 1/6$ is introduced to represent experimental decoherence. The trapped-ion realization of the Holstein and Fr\"ohlich models reproduces the desired spectral functions, demonstrating all the key polaron features, including single phonon excitation band, shift in the ground-state energy due to polaron binding, and single particle residue renormalization. Therefore low-energy polaron physics can be realized and probed using trapped-ion setups. 

\subsection{Long range hopping renormalization}

Here, we apply the displaced initial state protocol to extract effective hopping renormalization in the long-range case. The key distinction from NN hopping is that different $J_{ij}$ hopping elements get renormalized differently, since the phonon cloud overlap depends on the hopping distance (see Sec.~\ref{sec:Holstein_and_Frohlich_intro}). Generally, in solid-state systems, the coupling of modes is monotonically decreasing with distance, since it arises from (screened) Coulomb interaction between ions and electrons, resulting in the Holstein and Fr\"ohlich models. Therefore, we expect hopping between more distant sites to get renormalized more strongly. 

To avoid the effect of static site-dependent potentials that can arise in finite systems (see Sec.~\ref{sec:LF}) in our protocol implementation, we exploit the system's inversion symmetry and extract the effective hopping between inversion-symmetric sites $n \leftrightarrow 7-n$. Fig.~\ref{fig:effective_hopping_nn}a schematically illustrates the protocol: a spin is initialized at a site with a Lang-Firsov-displaced phonon cloud $\ket{i,\alpha^{\rm LF}_i}$ and then hops to the next site, adiabatically followed by the phonon cloud. This allows us to extract the effective hopping $J_{n,7-n}^{\rm eff}$. However, the presence of multiple hopping timescales complicates the extraction of effective hopping. Therefore, to directly extract the hopping elements, we extend the protocol and constrain the dynamics to the two sites of interest by setting all the $J_{i,j}$ elements except $J_{n,7-n}$ and $J_{7-n,n}$ to zero. This is possible in trapped-ion systems by illuminating only the ions of interest with the multi-frequency laser drives generating the spin--phonon interactions.

We show the extracted hopping renormalization $J_{ij}/J^{\rm eff}_{ij}$ as a function of coupling strength for the `ideal' and `trapped-ion' Holstein model with $\xi = 0.83$ and Fr\"ohlich model with $\alpha = 1.6$, for long-range hopping with $\alpha_J = 1.2$ in Figs.~\ref{fig:long-range_hol_vs_fro}(c) and \ref{fig:long-range_hol_vs_fro}(d), respectively. For both models, we observe that the more distant hopping elements, represented by $J_{25}$, get renormalized more strongly than the closer distant ones, represented by $J_{34}$, as suggested by the Lang-Firsov theory. Furthermore, for the same value of coupling $\Lambda$, we see that the effective mass is renormalized more strongly in the Holstein model. The agreement between ideal and trapped-ion realizations is better for the Fr\"ohlich model, since the trapped-ion systems naturally realize long-range interactions, given the long-range nature of the phonon modes. In the Holstein case, we see qualitative agreement with the ideal case and an increase of effective mass by a factor of two at $\Lambda\sim 4$.

We note a consequence of long-range hopping. Due to multiple possible hopping processes $i \to j $, the overall dynamics is sensitive to individual hopping renormalizations, which depend on the exact shape of the polaronic cloud, i.e., the spin dynamics are influenced by phonon correlation on the length scale given by the hopping range. Therefore, the analysis of individual hopping renormalizations may be used to characterize the shape of the polaron cloud, in conjunction with the \textit{polaron tomography}.  

\section{Experimental measurement protocols} \label{sec:protocols}

The primary means of measuring observables in trapped-ion platforms is through state-dependent fluorescence of the encoded qubit \cite{Noek13}. The two eigenstates of the qubit in the $\sigma_z$ basis $\{ \ket{\uparrow_z},\,\ket{\downarrow_z} \}$ are mapped to a pair of energy levels encoded in the ground, optical, and/or metastable states in trapped ions \cite{Allcock2021}.
The combination of site-resolved imaging and entangling operations between spins and collective bosons of trapped-ion Coulomb crystals enables the measurement of higher-order moments of both spin and bosonic observables, beyond single-spin magnetization $\{\langle\sigma^\phi_i\rangle\}$. This includes average bosonic displacement $\{\langle x_{\lambda}\rangle\}$ \cite{Gerritsma2010}, bosonic characteristic function \cite{fluhmann2020direct}, average bosonic occupation number $\{\langle a^{\dagger}_{\lambda}a_{\lambda}\rangle\}$ \cite{RasmussonOptimalSBC, leibfried2003quantum}, and multi-body spin correlations $\{\langle\sigma^\phi_i\sigma^{\phi'}_j\rangle\}$ \cite{Monroe2021}.
While the above protocols are sufficient to measure the properties of the normal modes or spins individually, such as measuring the phonons' characteristic function or extracting the effective polaronic mass from spin transport dynamics, measuring correlations between the spins and bosons is not straightforward.

In this section, we outline the key experimental protocols for measuring spin--phonon correlations developed in polarons and the displaced Green's function mentioned in Sec.~\ref{sec:LDOS+Tomography}. The protocols below focus on the case of the SLP mapping of the normal mode operators to local-phonon modes for simplicity, but they can be generalized to other L\"owdin orthogonalized mapping methods discussed in Sec.~\ref{sec:Engineering_electron-phonon_models}.

\subsection{Spin--phonon correlations}

\begin{figure}
    \centering
    \includegraphics[width=1.0\linewidth]{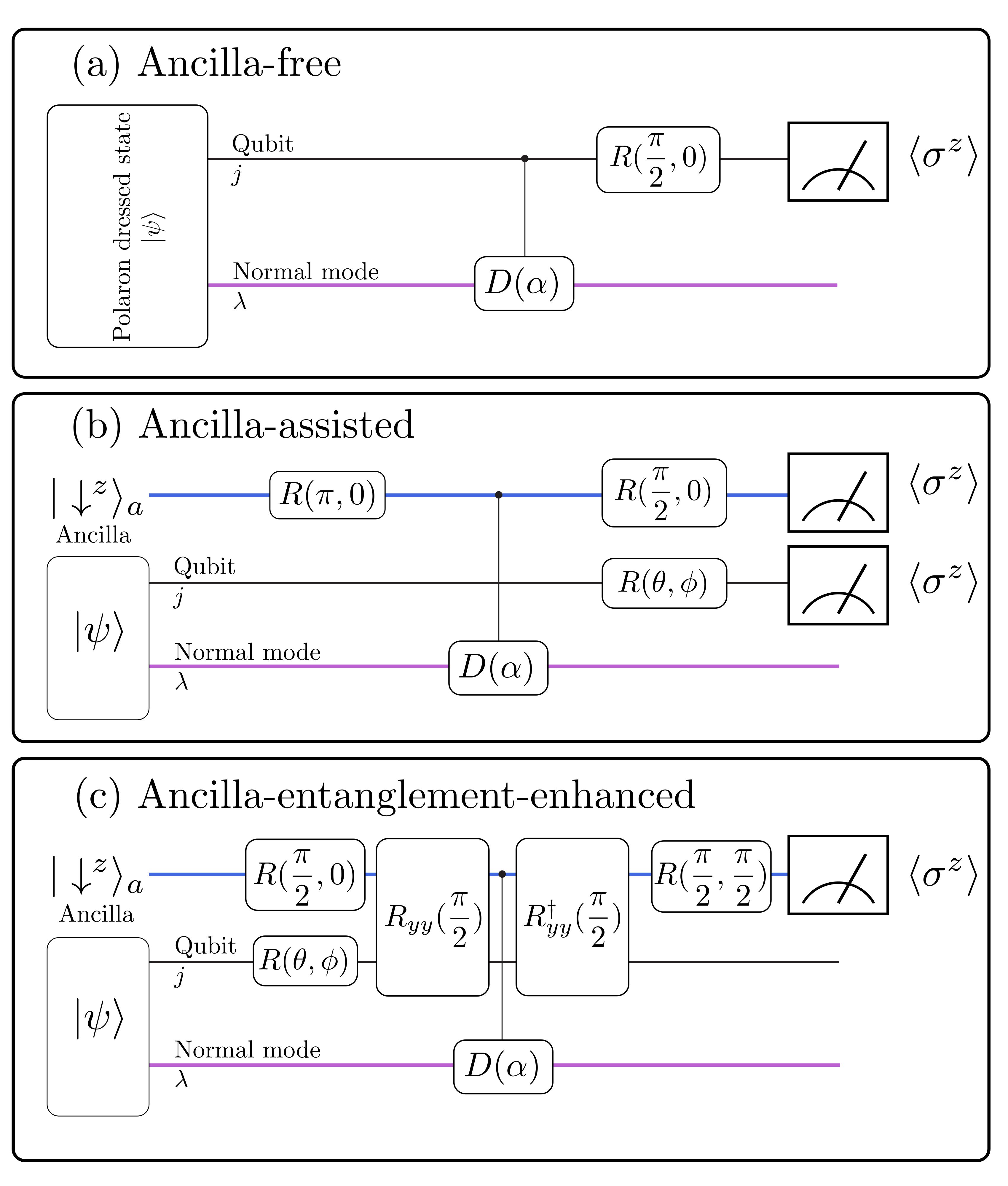}
    \caption{{\bf Quantum circuits for spin-phonon correlation measurements.}
    {\bf(a)} After forming the polaron dressed state, a data qubit $j$ can be first coupled weakly to the phonon quadrature of the target normal mode $\lambda$ via a spin-dependent force and then used to infer the correlation from the slope of the Pauli observable measured in the experiment basis $\sigma^z$.
    {\bf(b)} Using an ancilla allows background-free measurement of the spin-phonon correlation function. We measure the data qubit's spin state in the desired basis after $R(\theta,\phi)$ and use the ancilla to measure the phonon quadrature simultaneously.  
    {\bf(c)} Alternate to (b), entangling the ancilla to the data qubit, weakly coupling to the normal mode, and next reversing the entanglement developed between the ancilla and data qubit propagates the spin--phonon correlator observable purely to the ancilla.
    }
    \label{fig:spinphoncorr_measure_circuit}
\end{figure}

Spin--phonon correlator observables provide crucial information about the development of spin--phonon coherence during polaron formation, beyond what is available from simply measuring spin magnetization and bosonic quadratures.
If an entangling interaction of form $H_{\rm sp-O}\propto\sigma^\gamma_j\otimes \boldsymbol{O}$ can be realized for a given linear Hermitian operator $\boldsymbol{O}$ auxiliary to the spin degree of freedom at site $j$, then both the operator itself, as an observable, and its correlation with the spin can be measured. In this section, we outline techniques to measure $\braket{\sigma^\gamma_j\, \boldsymbol{O}}$ in the context of $\boldsymbol{O}=x_\lambda$, the displacement operator of normal mode $\lambda$ (see Fig.~\ref{fig:spinphoncorr_measure_circuit}). For a general L\"owdin-orthogonalized transformation, linear local mode phonon operators are a linear transformation of the normal mode phonon operators, for e.g. $x_m=\sum_\lambda V_{m\lambda} x_\lambda $. 

{\bf\emph{(i)}} \emph{Ancilla-free:} To measure the correlation $\langle \sigma^{\gamma}_j\,  x_\lambda\rangle$ without ancillas, we apply a weak coupling pulse to the data ion itself \cite{Gerritsma2010}. The unitary
\begin{equation}
U_{\lambda} = \exp(-ig\,\delta t\, \sigma^{\alpha}_j\,  x_\lambda),
\label{eq:weakdispacementprobe}
\end{equation}
where $g$ is the spin--phonon coupling strength. In the weak-coupling limit, where $\delta t \ll 1/g$, this probe pulse yields the expectation value $\langle\sigma^{\beta}_i\rangle = \langle\psi| U_{\lambda}^{\dagger} \sigma^{\beta}_i U_{\lambda} |\psi\rangle$ after preparing $\ket{\psi}$ using Eq.~\eqref{eq_tot_H_ions}. Expanding to first order in $\delta t$, we get
\begin{align}
\langle\sigma^{\beta}_j\rangle &\simeq \langle\psi| [1 + ig\delta t \sigma^{\alpha}_j \otimes x_{\lambda}] \sigma^{\beta}_j [1 - ig\delta t \sigma^{\alpha}_j \otimes x_{\lambda}] |\psi\rangle \nonumber \\
&\simeq \langle\psi| \sigma^{\beta}_j |\psi\rangle + ig\delta t \langle\psi| [\sigma^{\alpha}_j, \sigma^{\beta}_j] \otimes x_{\lambda} |\psi\rangle + O(\delta t^2) \nonumber \\
&= \langle\sigma^{\beta}_j\rangle_0 - 2g\delta t \epsilon^{\alpha\beta\gamma} \langle \sigma^{\gamma}_j x_{\lambda} \rangle,
\label{eq:ancillafree_sp}
\end{align}
where the target correlation is extracted from the slope in $-2g\delta t$, provided $\langle\sigma^{\beta}_i\rangle_0$ is measured independently in general, and $\epsilon^{\alpha\beta\gamma}=1$ is the Levi-Cevita symbol. However, because of the structure of Eq.~\eqref{eq_tot_H_ions} that inhibits multiple spin excitations in the $z$ basis, the spins are constrained to remain in an eigenstate of $\sigma^z$, which thereby suppresses the first term $\braket{\sigma^y_j}$ for $\alpha=x$ in Eq.~\eqref{eq:weakdispacementprobe}.
Other bosonic quadratures can be probed by tuning the motional phase of the coupling pulse \cite{fluhmann2020direct}. In a general L\"owdin-orthogonalized map, measuring the spin-local phonon correlation $\braket{\sigma_j^\alpha x_m}$ uses $U_m$ as the weak coupling pulse, which is a composite pulse sequence of the normal mode coupling pulse $U_\lambda$, following
\begin{align}
\label{eq:probelocalmodes}
U_m=&\Pi^{M}_{\lambda} (U_{\lambda})^{V_{m\lambda}}
\nonumber\\=\,&e^{-ig\,\delta t\, \sigma^{\alpha}_j\, \sum^M_{\lambda}(V_{m\lambda} x_\lambda)}
\nonumber \\=\,&e^{-ig\,\delta t\, \sigma^{\alpha}_j\, x_m},
\end{align}
implemented either in series or in parallel. This transformed pulse $U_m$ can be readily substituted for $U_\lambda$ in the other protocols discussed in this section.

{\bf\emph{(ii)}} \emph{Ancilla-assisted:} To yield a better signal-to-noise ratio by eliminating the residual magnetization in Eq.~\eqref{eq:ancillafree_sp}, we can divide the tasks of measuring the bosonic quadrature and the spin to an ancilla ion and a data ion, respectively. 
We consider the observable of interest $\braket{\sigma^{\gamma'}_jx_\lambda}$, where spin $j$ belongs to the system $\ket{\psi}$ and ancilla spin $a$ is used for the bosonic tomography akin to the ancilla-free approach.
As shown in Fig.~\ref{fig:spinphoncorr_measure_circuit}(b), we start from $\ket{\psi}\otimes\ket{\downarrow^z_a}$, and the ancilla is prepared in the eigenstate $\ket{\uparrow^z_a}$ using a single-qubit rotation.
Next, the weak-coupling pulse, $U^x_\lambda = \exp(-ig\,\delta t\, \sigma^{x}_a\,  x_\lambda)$, is applied between the ancilla ion and the normal mode.
In parallel, the data qubit can be prepared in the desired basis after measurement using single-qubit rotations. This results in
\begin{align}
\langle\sigma^{\gamma'}_j\,\sigma^{y}_a\rangle &\simeq
    \langle
    \sigma^{\gamma'}_j\, \big(
    \sigma^{y}_a - 2g\,\delta t\,  \sigma^{z}_a \,x_\lambda \big )\rangle \nonumber\\
    &=- 2g\delta t\braket{\sigma^{\gamma'}_j\, x_\lambda }
    ,
    \label{eq:ancillaassisted_sp}
\end{align}
where the slope of the correlated measurement of the two ions with respect to $-2g\delta t$ reveals the spin--phonon correlation, and the residual magnetization is removed by preparing the ancilla in an eigenstate of $\sigma^z_a$.

{\bf\emph{(iii)}} \emph{Ancilla-entanglement-enhanced:}

As an alternative ancilla-assisted protocol, the short ancilla-phonon probe can be dressed by sandwiching it between a phase-aligned Mølmer-Sørensen (MS) gate ~\cite{Molmer1999} and its inverse, namely $U_{\rm MS}^\dagger\left(\frac{\pi}{2}\right) U_{a}(\phi,\psi,\delta t)\,U_{\rm MS}\left(\frac{\pi}{2}\right)$. For the initial ancilla state $\ket{\uparrow^{\phi+\pi/2}}_a$, the MS-gate pair maps the system spin component $\sigma^{\phi+\pi/2}_{j}$ onto the ancilla readout channel associated with $\sigma^z_{a}$,
\begin{align}
    &U_{\rm MS}^\dagger\left(\frac{\pi}{2}\right) U_{a}(\phi,\psi,\delta t)\,U_{\rm MS}\left(\frac{\pi}{2}\right) \\&=\exp\left(i\,{g\delta t}\,\sigma^{\phi+\pi/2}_{j}\sigma^z_{a}x_{\lambda}\right),
\end{align}
where
\begin{equation}
    U_{\rm MS}\left(\frac{\pi}{2}\right)=\exp\left(-i\frac{\pi}{4}\sigma^{\phi+\pi/2}_{j}\sigma^{\phi+\pi/2}_{a}\right).
\end{equation}
 The implementation proceeds as follows. As in the previous ancilla-assisted protocol, the ancilla is initialized, such that the total system state is $\ket{\Psi_0}=\ket{\psi}_{s,\rm ph}\otimes\ket{\uparrow^{\phi+\pi/2}}_a$, where $\ket{\psi}_{s,\rm ph}$ is the system-phonon state of interest. After the first MS gate, the state of the total system becomes
\begin{equation}
    \ket{\Psi_1}=\left[\exp\left(-i\frac{\pi}{4}\sigma^{\phi+\pi/2}_{j}\right)\ket{\psi}_{s,\rm ph}\right]\otimes\ket{\uparrow^{\phi+\pi/2}}_a.
\end{equation}
Next, the coupling drive $U_{a}(\phi,\psi,\delta t)$ is applied, resulting in
\begin{equation}
    \ket{\Psi_2}=\exp\!\left(-i\,{g\, \delta t}\,\sigma^\phi_{a} x_{\lambda}\right)\ket{\Psi_1}.
\end{equation}
In the weak-coupling limit given by $g\,\delta t \ll 1$, this can be expanded to first order as
\begin{equation}
    \ket{\Psi_2}\approx\ket{\Psi_1}-i\,{g\,\delta t}\,\sigma^{\phi}_{a}x_{\lambda}\ket{\Psi_1}.
\end{equation}
Since $U_{\rm MS}^\dagger\left(\frac{\pi}{2}\right)\ket{\Psi_1}=\ket{\Psi_0}=\ket{\psi}_{s,\rm ph}\otimes\ket{\uparrow^{\phi+\pi/2}}_a$, $U_{\rm MS}^\dagger\left(\frac{\pi}{2}\right) \sigma^{\phi}_{a}\, U_{\rm MS}\left(\frac{\pi}{2}\right)= - \sigma^{\phi+\pi/2}_{j}\sigma^z_{a}$, and $\sigma^z_{a}\ket{\uparrow^{\phi+\pi/2}}_a=\ket{\downarrow^{\phi+\pi/2}}_a$, applying an inverse MS pulse then yields
\begin{align}
    \ket{\Psi_3} &\approx \ket{\psi}_{s,\rm ph}\otimes\ket{\uparrow^{\phi+\pi/2}}_a\\
    &+i\,{g\, \delta t}\,\sigma^{\phi+\pi/2}_{j}x_{\lambda}\ket{\psi}_{s,\rm ph}\otimes\ket{\downarrow^{\phi+\pi/2}}_a.
\end{align}
By using $\sigma^{\phi}\ket{\uparrow,\downarrow^{\phi+\pi/2}}=\pm i\ket{\downarrow,\uparrow^{\phi+\pi/2}}$, $\bra{\Psi_0}\sigma^{\phi}_{a}\ket{\Psi_0}=0$. Correspondingly, to the first order in $g\delta t$, we get
\begin{equation}
    \bra{\Psi_3}\sigma^{\phi}_{a}\ket{\Psi_3}\approx\,2{g\, \delta t}\,\langle\sigma^{\phi+\pi/2}_{j} x_{\lambda}\rangle.
\end{equation}
A final analysis pulse $R^{\phi+\pi/2}_{a}\left(\frac{\pi}{2}\right)$ is applied to map the ancilla observable $\sigma^{\phi}_{a}$ onto $\sigma^z_{a}$ for fluorescence readout
\begin{equation}
    \braket{\sigma^z_{a}}_{\rm final}=\bra{\Psi_3}\sigma^{\phi}_{a}\ket{\Psi_3}\approx
    2\,{g\, \delta t}\,\langle\sigma^{\phi+\pi/2}_{j} x_{\lambda}\rangle.
\end{equation}
This protocol also provides a zero-baseline ancilla-fluorescence signal for the target spin-phonon correlator, like the previous scheme. We assume the MS gates use radial normal modes orthogonal to those used for the spin-phonon displacement operations, which helps to reduce gate infidelity from residual spin-phonon correlation. Ideally, this method is the most favorable of the three since the detection error is limited to that of the ancilla ion only. In practice, however, imperfections in the MS gate may reduce or even eliminate the advantage of its zero-baseline readout.

The techniques presented here can be further extended to mapping the bosonic characteristic function of multiple phonon modes to observe higher-order correlators \cite{fluhmann2020direct}. A notable concern is the heating of the motional modes during fluorescence detection~\cite{rasmusson2024measurement}. This can be mitigated by post-selecting dark state measurements and alternating spin-state assignments between measurement runs to minimize phonon scattering~\cite{Gerritsma2010}.

\subsection{Displaced-Fock Green's function}\label{sec_dfock_GF}

The displaced Green's function encodes the full spectral information of the system and provides the key input for computing spectral functions and local density of states as discussed in Secs.~\ref{sec:LDOS+Tomography} and \ref{sec:Long-range_hopping_polaron}. We generalize the displaced Green's function in Eq.~\eqref{eq:displacedGF} and describe it in terms of a displaced-Fock basis as
\begin{align}
    \nonumber G^{ {\alpha_F,\beta_{F'}}}&(j,l,t) =\\& -i\,\langle\psi^{0}_{\rm sys}|\sigma^{-}_l F'\,D( {\beta})\,  e^{-iHt} D( {\alpha})\,F\,\sigma^{+}_j\ket{\psi^{0}_{\rm sys}},
    \label{eq:displacedFockGF}
\end{align}
using a unitary operator $F$ generated by an anti-Jaynes-Cummings Hamiltonian to introduce multiple single phonon excitations before the initial and after the final displacements. Here, the net initial and final displacements of the phonons are $D({\alpha})=\Pi_{\lambda=1}^{M}D(\alpha_\lambda)$ and $D({\beta})=\Pi_{\lambda=1}^{M}D(\beta_\lambda)$, respectively, and $\ket{\psi^{0}_{\rm sys}}=\Pi^{N}_{i=1}\ket{\downarrow^z}_i\otimes\ket{ {0_{\rm ph}}}$ denotes the system's vacuum state in spin and phonon excitations. 
In general, preparing a superposition of displaced or displaced-Fock states $\ket{\alpha_F}$ and $\ket{\beta_F}$ that have considerable overlap with the ground state of the Hamiltonian provides the most information about the polaronic ground state.

\begin{figure}
    \centering
    \includegraphics[width=1.0\linewidth]{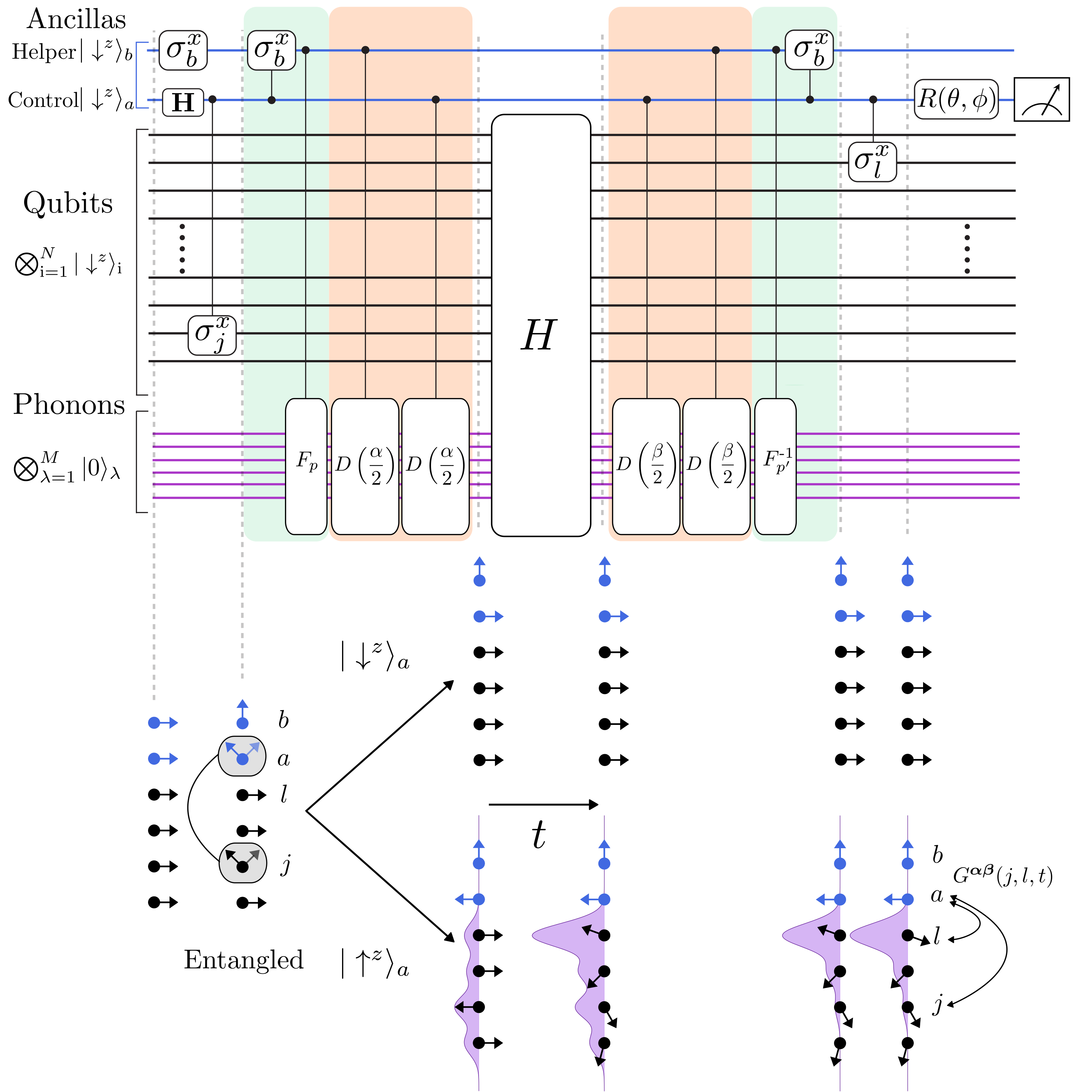}
    \caption{\textbf{Displaced Green's function measurement protocol.}
    The Green's function is measured between spins at sites $j$ and $l$ under a controlled evolution by the Hamiltonian depending on the state of the control ancilla ion. After conditionally exciting spin $j$, the control and helper ancilla ions first help to create multiple phonon Fock excitations (first green chapter) and displace the phonons (first orange chapter) to prepare close to the desired polaronic state. Next, the Hamiltonian evolved state is projected onto displaced-Fock states by first displacing the phonons (second orange chapter) and then performing Fock-state operations (second green chapter). At the end of the circuit, a single-qubit rotation pulse $R(\theta,\phi)$ on the control ancilla enables the extraction of the real and imaginary parts of $G^{ {\alpha_F\beta_{F'}}}(j,l,t)$. A pictorial representation of the system's evolution in the spin states and local phonon displacements (purple envelopes) following each set of operations is shown below the circuit.
    }
    \label{fig:GF_measure_circuit}
\end{figure}

The ancilla-assisted protocol to measure $G^{ {\alpha \beta}}(j,l,t)$, as elucidated in Fig.~\ref{fig:GF_measure_circuit}, proceeds through the following steps. We initialize the system, control ancilla, and helper ancilla in their ground state $\ket{\Psi_0} = \ket{\downarrow^z}_b \otimes\ket{\downarrow^z}_a \otimes \ket{\psi^{0}_{\rm sys}}$. We then apply a Hadamard gate on the control ancilla and a $\sigma^x$ rotation on the helper ancilla, resulting in
\begin{equation}
\ket{\Psi_1} = \ket{\uparrow^z}_b \otimes\frac{1}{\sqrt{2}} \left(\ket{\downarrow^z}_a + \ket{\uparrow^z}_a\right) \otimes\ket{\psi^{0}_{\rm sys}}.
\end{equation}
Next, we apply ancilla-controlled $\sigma_j^{x}$ (CNOT) gate, creating the state
\begin{equation}
\ket{\Psi_2} = \ket{\uparrow^z}_b \otimes\frac{1}{\sqrt{2}}
\big(
\ket{\downarrow^z}_a \otimes \ket{\psi^{0}_{\rm sys}} + \ket{\uparrow^z}_a\otimes \sigma_j^{x} \ket{\psi^{0}_{\rm sys}}
\big).
\end{equation}

For the Hamiltonian discussed here, the CNOT gate serves to controllably excite spin $j$ similar to a `controlled-$(\sigma^{\pm})$' gate, which could have been useful to measure the Green's function, but is non-unitary. To prepare a product state of Fock excitations conditioned on the control ancilla, we first entangle the helper and control ancillas with another CNOT gate, leading to 
\begin{align}
\nonumber \ket{\Psi_3} =& \frac{1}{\sqrt{2}}
\big(
\ket{\uparrow^z}_b\otimes\ket{\downarrow^z}_a \otimes \ket{\psi^{0}_{\rm sys}} +\\
&\ket{\downarrow^z}_b\otimes\ket{\uparrow^z}_a\otimes \sigma_j^{x} \ket{\psi^{0}_{\rm sys}}
\big).
\end{align}

A single Fock excitation of mode $p$, where $\{p\}\subseteq\{\lambda\}$, can be generated using $F_{p}=\text{exp}(-i\, \pi/2\, [\sigma^{-}_{b} a_p e^{-i\phi_p} +\sigma^{+}_{b} a_p^\dagger e^{i\phi_p} ])$~\cite{leibfried2003quantum}. 
A repeated implementation of the combined CNOT and $F_{p}$ operations (first green chapter in Fig.~\ref{fig:GF_measure_circuit}) allows multiple $p$ modes to have a single phonon excitation. Similarly, higher Fock excitation numbers can also be generated, provided the $\pi$-pulse-time accounts for the Debye-Waller factors \cite{leibfried2003quantum}. The complete unitary operation is $F=\Pi_p\,(F_p \,U^{\text{CNOT}}_{ab})$, leading to the state
\begin{equation}
\ket{\Psi_4} = \ket{\uparrow^z}_b \otimes\frac{1}{\sqrt{2}}
\big(
\ket{\downarrow^z}_a \otimes \ket{\psi^{0}_{\rm sys}} + \ket{\uparrow^z}_a\otimes
\sigma_j^{x} \ket{\psi^{1_F}_{\rm sys}}
\big),
\end{equation}
where $ \ket{\psi^{F}_{\rm sys}}=F\ket{\psi^{0}_{\rm sys}}=\Pi_p (-ie^{i\phi_p})^{|p|} \,a^\dagger_p\ket{\psi^{0}_{\rm sys}}$, and the phase $(-ie^{i\phi_p})^p$ can be nulled by tuning the laser phase $\phi_p$.
Optionally, omitting $F$ throughout this protocol yields the Green's function in the displaced basis instead of the displaced-Fock basis.
Next, a displacement pulse $D( {\alpha})$ of form $\sim\sigma^z(a_\lambda e^{-i\phi_\lambda}+a_\lambda^\dagger e^{i\phi_\lambda})$ is applied on the helper and the control ancillas sequentially to project the normal modes closer to the polaronic ground state (first orange chapter in Fig.~\ref{fig:GF_measure_circuit}). The helper ancilla first displaces by $\ket{+ {\alpha}/2}$, and next the control ancilla conditionally displaces the normal modes to $\ket{\pm {\alpha}/2}$ depending on its spin state, resulting in 
\begin{align}
\ket{\Psi_5}  =&\ket{\uparrow^z}_b \otimes \frac{1}{\sqrt{2}}
\big(
\ket{\downarrow^z}_a \otimes\ket{\psi^{0}_{\rm sys}}  +\ket{\uparrow^z}_a\otimes\, \ket{\psi^{ {\alpha_F}}_{j}}
\big),
\end{align}
where $\ket{\psi^{ {\alpha_F}}_{j}}=D( {\alpha})\, \sigma_j^{x} \ket{\psi^{1_F}_{\rm sys}}$. 

The system then evolves under $U_H(t) = e^{-iHt}$ primarily for $\ket{\uparrow^z}_a$ due to a single spin excitation in the system, while the other branch $\ket{\downarrow^z}_a$ evolves with the phase $e^{i Et}$, where $E$ represents the net $\sigma^z$ rotation frequency from the Hamiltonian. We obtain
\begin{equation}
\begin{aligned}
\ket{\Psi_6}  =&\ket{\uparrow^z}_b \otimes \frac{1}{\sqrt{2}}
\big(
\ket{\downarrow^z}_a \otimes e^{iEt} \ket{\psi^{0}_{\rm sys}} \\& +\ket{\uparrow^z}_a\otimes\, U_H(t)\,\ket{\psi^{ {\alpha_F}}_{j}}
\big)
\end{aligned}
\end{equation}

To obtain the projection of the evolved state into a displaced-Fock state, we first apply another displacement $D(\beta)$ using a similar, but reversed, protocol as $D(\alpha)$ (second orange chapter in Fig.~\ref{fig:GF_measure_circuit}). While $\beta$ can be arbitrary, we can obtain the corresponding dLDOS when $\beta=-\alpha$ and $j=l$, which reverses the initial displacements.
This projection is completed by applying $F'=\Pi_{p'} ( \,(U^\dagger)^{\text{CNOT}}_{ab}F_{p'}^{-1})$, where $\{p'\}\subseteq{\{\lambda\}}$, in a reversed sequence with respect to the initial Fock excitation operation (second green chapter in Fig.~\ref{fig:GF_measure_circuit}). To obtain the dLDOS, the Fock excitation operation at the beginning of this protocol is the inverse of that at the end $F^{-1}_{p'}=(F_{p})^{-1}$, where $p=p'$. In general, the state becomes
\begin{equation}
\begin{aligned}
\ket{\Psi_7} =& \ket{\uparrow^z}_b \otimes\frac{1}{\sqrt{2}}
\big(
\ket{\downarrow^z}_a \otimes\, e^{iEt} \ket{\psi^{0}_{\rm sys}} 
\\& + \ket{\uparrow^z}_a\otimes F'\,D( {\beta})\,U_H(t) \,\ket{\psi^{ {\alpha_F}}_{j}}
\big).
\end{aligned}
\end{equation}

We then apply $\sigma_l^{x}$ conditioned on the control ancilla, yielding
\begin{equation}
\begin{aligned}
\ket{\Psi_8} =& \ket{\uparrow^z}_b \otimes\frac{1}{\sqrt{2}}
\big(
\ket{\downarrow^z}_a \otimes\, e^{iEt} \ket{\psi^{0}_{\rm sys}} 
\\& + \ket{\uparrow^z}_a\otimes \sigma_l^{x}\,F'\,D( {\beta})\,U_H(t) \,\ket{\psi^{ {\alpha_F}}_{j}}
\big).
\end{aligned}
\end{equation}
The initial and final CNOT gates can be realized using the same radial modes that are driven to realize the spin-hopping Hamiltonian. Residual spin--phonon entanglement at the end of the evolution, which might limit the second gate fidelity, can be mitigated by setting the evolution time to be a multiple of the detuning frequency $\delta_s$ used to realize the spin-hopping Hamiltonian, $t=n\, 2\pi/\delta_s$.
The reduced density matrix of the control ancilla after this sequence is
\begin{align}
\rho_a &= \frac{1}{2}
\begin{pmatrix}
1 & e^{-iEt}\, C \\
e^{iEt} \,C^* & 1
\end{pmatrix},
\end{align}
where $C=\langle{\uparrow^z_b,\psi^{0}_{\rm sys}} \ket{\uparrow^z_b, \psi ^{ {\alpha_F,\beta_{F'}}}_{j,l}(t)}$, and  $\ket{\psi ^{ {\alpha_F,\beta_{F'}}}_{j,l}(t)}=\sigma_l^{x}\,F'\,D( {\beta})\,U_H(t) \,D( {\alpha})\,F\,\sigma_j^{x} \ket{\psi^{0}_{\rm sys}}$.
The Green's function of interest is $G^{ {\alpha_F \beta_{F'}}}(j,l,t)=-i\,C$.
Finally, we measure in the computational $z$ basis after applying Pauli operators on the control ancilla to extract the real and imaginary components of the Green's function
\begin{align}
\langle \sigma^x_a \rangle &= \text{Tr}(\rho_a \sigma^x_a) = \Re{i e^{-iEt}G^{ {\alpha_F,\beta_{F'}}}(j,l,t)}, \\
\langle \sigma^y_a \rangle &= \text{Tr}(\rho_a \sigma^y_a) = \Im{-i e^{-iEt}G^{{\alpha_F,\beta_{F'}}}(j,l,t)},
\\
&G^{{\alpha_F,\beta_{F'}}}(j,l,t)=-i\, e^{iEt} (\braket{\sigma^x_{a}}-i\braket{\sigma^y_{a}}).
\label{eq:GFinPauliexpressions}
\end{align}
\vspace{-0.5cm}

By repeating this protocol for different sites $j,l$ and evolution times $t$, the full real-time Green's function is reconstructed, enabling computation of spectral functions and dLDOS of the polaronic state via Fourier transformation. If the global phase $Et$ is not accounted while evaluating Eq.~\eqref{eq:GFinPauliexpressions}, the measured Green's function spectrum only shifts by a constant from $G^{ {\alpha_F,\beta_{F'}}}(j,l,\nu)\rightarrow G^{ {\alpha_F,\beta_{F'}}}(j,l,\nu-E)$ and can be calibrated out.

\section{Discussions and Outlook}
\label{sec:Discussion}

In this work, we provide a blueprint for a trapped-ion architecture in which the range and strength of spin--phonon couplings and spin hopping can be independently controlled through near-resonant and off-resonant driving of transverse phonon modes in ion crystals, respectively. We developed a dynamical measurement framework that accesses polaron properties without requiring adiabatic preparation of the many-body ground state. Displaced Green's functions, implemented through spin-dependent phonon displacements, restore spectral overlap with strongly dressed polaronic quasiparticles and complement conventional single-particle spectroscopy in the strong-coupling regime. Combined with Lang--Firsov-motivated initial states, these correlators and real-time observables provide access to equilibrium quantities, such as quasiparticle energies, renormalized masses, binding energies, and quasiparticle weights.  Spatially resolved displaced Green's functions, together with spin, phonon, and spin--phonon correlators, further resolve the structure of the dressing cloud, providing access to polaron tomography. 
These results establish a framework in which the formation, dynamics, and spatial structure of polarons can be characterized on a single programmable platform, with the spin--phonon coupling profile and initial phonon displacement serving as independent controls. More broadly, this level of control makes the phonon dressing cloud itself programmable: rather than merely reproducing prescribed Holstein- or Fr\"ohlich-type interactions, one may optimize the spatial spin--phonon coupling matrix and the initial phonon displacement to realize a desired spectral or transport response, providing a route toward the inverse design of polaronic properties relevant to quantum materials.

Extending the simulator beyond the single-excitation sector by adding more spin excitations to the initial state would move the platform from single-polaron physics to the few- and ultimately many-carrier regime. The two-carrier problem provides a particularly direct next step, where it would allow one to determine whether phonon-mediated interactions, together with the long-range spin--spin interactions responsible for carrier hopping, are sufficient to bind two dressed carriers into a bipolaron~\cite{PhysRevB.104.214309, PhysRevX.13.031017}. Bipolaron formation could be investigated using spatial correlations between the two carriers and dynamical two-particle correlators. The spin-density, phonon-density, and bare Green's function observables introduced here remain applicable in the multi-carrier regime, while identifying a bipolaron would require an appropriately constructed displaced two-particle Green's function or a pair propagator, which we leave for future work.

Increasing the carrier density further would provide access to genuinely many-body phenomena, in which the competition among carrier mobility, spin correlations, and phonon-mediated interactions can lead to collective behavior beyond isolated polarons or bipolarons. In particular, this architecture could be used to investigate the emergence of pairing correlations, charge-density order, and phase separation as functions of carrier density and interaction strength~\cite{Knoerzer2022, Bhattacharya2021PhononInducedPairing}. The spin and spin--phonon observables available in the present scheme already provide access to spatial density correlations and lattice displacements in this regime. Establishing superconducting or other coherent pairing phases, however, requires additional information on off-diagonal and long-range pair correlations, which remains an open problem in the current trapped-ion setting. This progression from single-particle dressing to few-body binding and ultimately to collective many-body order provides a natural route for extending the present simulator toward strongly correlated polaronic matter.

A further extension of the platform would be to exploit ions with three or more individually addressable internal states, providing local qudit degrees of freedom beyond the two-level systems used here~\cite{Hrmo2023}. In particular, a three-level encoding directly realizes the local Hilbert space of a doped bosonic $t$-$J$ model, with the three states representing spin-$\uparrow$, spin-$\downarrow$, and a hole. Combined with the phonon modes of the ion crystal, this would provide a direct route to studying doped bosonic $t$-$J$ models with dynamical lattice degrees of freedom, including stripe formation and kinetic magnetism~\cite{Homeier2024, Harris2026}. Coupling the holes to phonons would then provide a controlled way to investigate how lattice dressing renormalizes hopping and modifies kinetic frustration, with implications for hole pairing and collective behavior. 

More generally, the protocols designed here can also be applied to site-resolved two-dimensional crystals \cite{Kiesenhofer2023}. Extending this encoding to two-dimensional trapped-ion crystals would make it possible to study these models on approximately triangular geometries, where hole motion is kinetically frustrated by the non-bipartite spin background, connecting naturally to the physics of triangular Moiré Hubbard systems and spin polarons~\cite{Schloemer2024, Tao2024, Prichard2024}. Simulating polaron models on a two-dimensional triangular lattice would also enable the investigation of anisotropic polaron dressing and direction-dependent effective masses, as well as two-dimensional $t-J$ models  \cite{Qiao2025} in the presence of phonons. However, the denser vibrational spectrum will require a tailored multi-frequency optimization to select specific site-phonon couplings~\cite{Guo2024site-resolved, Qiao2024tunable, Hou2024}. 

Beyond the parameter regimes considered here, where thermal effects are negligible, the platform gives access to dynamics at temperatures comparable to the spin and phonon energy scales~\cite{So2026}.  At finite temperature, one could quantify how carrier--phonon correlations and many-body signatures, such as pairing or charge ordering, evolve with temperature, including the temperature dependence of correlation lengths, pair-binding energies, and dynamical response functions. Recent trapped-ion experiments have demonstrated controlled preparation of motional states at tunable temperatures, providing a route toward finite-temperature spin--boson simulations~\cite{Sun2025, So2026}. 

Trapped-ion platforms also allow for engineered dissipation~\cite{so2024trappedion,so2025, So2026} and control over effective phonon spectra and dispersion, going beyond the optical phonons considered here. Controlled dissipation offers a complementary means of modifying relaxation and nonequilibrium dynamics, enabling engineered relaxation pathways, state preparation, and nonequilibrium steady states~\cite{Barreiro2011, Reiter2021, Zhu2025}. The vibrational environment could likewise be tuned from stiff optical modes towards soft, more dispersive acoustic phonons, allowing the roles of phonon dispersion and spatially extended dressing to be investigated directly. Introducing controlled anharmonicity would enable phonon--phonon interactions and occupation-dependent vibrational frequencies, allowing the exploration of the interplay between nonlinear lattice dynamics and carrier dressing beyond the harmonic limit. Together, these capabilities would allow the roles of temperature, dissipation, and phonon spectral structure in carrier transport and energy relaxation to be resolved, with relevance to vibrationally assisted excitation and energy transfer in molecular aggregates and light-harvesting systems~\cite{Fassioli2014}.

\begin{acknowledgments}
U.B., F.M., and E.D. acknowledge financial support from the Swiss National Science Foundation
(SNSF) through project No.~200021\_212899, the SNSF Sinergia grant CRSII–222792, the Swiss State Secretariat for Education, Research and Innovation (SERI) under contract No.~UeM019-1, and NCCR SPIN, a National Center of Competence in Research funded by the Swiss National Science Foundation (grant No.~225153). U.B. is also grateful for the financial support of the IBM Quantum Researcher Program.  U.B. and F.M. would also like to acknowledge useful discussions with Jungsang Kim, O. Malyshev, S. Chattopadhyay, and L. Skolc.

G.P. acknowledges support from the Welch Foundation Award (grant no. C-2154), the NSF CAREER Award (grant no. PHY-2144910), the NSF NRT Award (grant no. 2346014), and the Office of Naval Research (grant no. N00014-23-1-2665). We acknowledge that this material is based on work supported by the U.S Department of Energy, Office of Science, Office of Nuclear Physics under the Early Career Award (grant no. DE-SC0023806). 

AI statement: We acknowledge the use of generative AI tools for grammatical correction, language and concept refinement, and auxiliary code in this work. All results were independently verified by the authors, who take full responsibility for the manuscript.
\end{acknowledgments}

\appendix

\section{Derivation of the spin--phonon Hamiltonian }\label{sec:app-deriving_Hamiltonian}

We consider a one-dimensional chain of $N$ ions confined in a harmonic Paul trap with site-resolved coherent control of the drive frequency, amplitude, and phase enabled by a multi-channel acousto-optic modulator, similar to setups in Refs.~\cite{Debnath_2016, Feng2023}. 

\subsection{Trapped-ion Hamiltonian }
We begin from the interaction-picture spin--motion Hamiltonian~\cite{Monroe2021,pagano2025varenna} for $N$ ions, labeled by $i$, coupled to $N$ collective normal modes of a relevant radial motional branch, labeled by $m$. The interaction is generated by $M\leq N$ bichromatic laser fields, labeled by $k$, with symmetric detunings $\mu_k$ from the qubit resonance and site-dependent spin and motional phases $\phi_{ik}$ and $\psi_{ik}$, respectively. In units of $\hbar=1$, the Hamiltonian reads
\begin{equation}
\label{eq_H_ions}
\begin{split}
H_I=&\sum_{i=1}^N \sum_{m=1}^N \sum_{k=1}^M \eta_{im} \Omega_{ik} \cos(\mu_k t+\psi_{ik})\\ & \times\left(a^\dagger_m e^{i\nu_m t} + a_m e^{-i\nu_m t}\right) \sigma^{\phi_{ik}}_{i}.
\end{split}
\end{equation}
Here, the coefficient $\eta_{im}=\eta_m \xi_{im}$ is the Lamb-Dicke parameter that weights the participation of the $i$-th ion on the $m$-th normal mode with motional mode frequency $\nu_m$. Its constituents are $\eta_m=\Delta k\sqrt{1/2m_{\rm ion}\nu_{m}}$, where $\Delta k$ is the photon momentum on the ions with mass $m_{\rm ion}$, and $\xi_{im}$, which is the orthogonal eigenvector matrix of the potential given by the harmonic trap and Coulomb interaction among the ions. The spin operators are $\sigma^{\phi_{ik}}_i=\sigma^x_i \cos(\phi_{ik}) + \sigma^y_i \sin(\phi_{ik})$ acting on the $i$-th ion. 

{\bf\emph{(i)}} \emph{Far-detuned drives for spin hopping:}
We first consider a radial motional branch driven sufficiently far
from all of its normal-mode sidebands, $|\mu_k-\nu_m|\gg \eta_{im}\Omega_{ik}$ $\forall i,m,k$. In this regime, the motional modes are only virtually excited, and the Magnus expansion of Eq.~\eqref{eq_H_ions} is dominated by the second-order term that gives rise to a pure two-body spin-spin Hamiltonian $H = \sum_{i<j} J_{ij}\sigma^x_i\sigma^x_j$ for equal spin phases $\phi_{ik}=0$. The full expression of the interaction matrix is \cite{Korenblit2012}
\begin{equation}
    J_{ij}=\omega_{\rm rec} \sum_{k=1}^M \sum_{m=1}^N \Omega_{ik} \Omega_{jk} \frac{\xi_{im} \xi_{jm}}{\mu_k^2 - \nu_m^2}\cos(\psi_{ik}-\psi_{jk}),
    \label{eq:Jij}
\end{equation}
where $\omega_{\rm rec}= \Delta k^2/2 m_\text{ion}$ is the photon recoil frequency.

For a single, far-detuned bichromatic drive ($M=1$) with $\mu_k=\mu$, $\Omega_{ik}=\Omega_i$, and $|\mu-\nu_m|\gg \eta_{im}\Omega_{i}$ $\forall i,m$ (see Fig.~\ref{fig:architecture}a), the interaction matrix is simplified to
\begin{equation}
    J_{ij}=\Omega_i \Omega_j \omega_{\rm rec} \sum_{m=1}^N \frac{\xi_{im} \xi_{jm}}{\mu^2 - \nu_m^2}\approx \frac{J_0}{|i-j|^\alpha},
    \label{eq_Jij}
\end{equation}
where $0.5 < \alpha < 1.8$ can be tuned over experimentally relevant $\mu$~\cite{Monroe2021}. The nearest-neighbor coupling $J_0$ and the exponent $\alpha$ are obtained from a best fit of the interaction matrix to a power law profile.

In the presence of a strong transverse field $B \gg J_0$, applied perpendicular to the spin--spin rotation axis associated with the Ising XX interaction above, the effective Hamiltonian becomes an exchange (XY) spin Hamiltonian that conserves the number of spin excitations:
\begin{equation}
\label{eq_XY}
H_{\rm hop}=\sum_{ij}J_{ij}(\sigma^+_i\sigma^-_j+\sigma^-_i\sigma^+_j).
\end{equation}

{\bf\emph{(ii)}} \emph{Near-resonant drives for spin--phonon coupling:}
We next consider the orthogonal radial motional branch used as the dynamical phonon sector. Each bichromatic tone $k$ is now chosen close to one normal mode, which we label by the same index $k$, with $\delta_k\equiv\mu_k-\nu_k$. The addressed mode is spectrally resolved from all other modes when $|\delta_k|,\; |g_{ik}|\ll|\mu_k-\nu_m|,\;m\neq k$, where $g_{ik}\equiv\eta_{ik}\Omega_{ik}$. For a common spin phase $\phi_{ik}=\phi$, retaining the near-resonant first-order spin--motion coupling and transforming to the frame rotating at the corresponding drive detunings (see Supplemental Material) yields
\begin{equation}
\begin{split}
\label{eq_spin_phonon}
H_{\rm rot}=&\sum_{i=1}^N\sum_{k=1}^M \frac{g_{ik}}{2} (a^\dagger_k e^{-i\psi_{ik}} + a_k e^{i\psi_{ik}})\sigma^\phi_i \\ &- \sum_{k=1}^M \delta_k a^\dagger_k a_k + H_{\rm cr},
\end{split}
\end{equation}
where $H_{\rm cr}=\sum_{i<j}J^{\rm cr}_{ij}\sigma_i^\phi\sigma_j^\phi$ is the residual crosstalk from each beatnote's off-resonant coupling to the other $N-1$ motional modes, with coupling strengths
\begin{equation}
\begin{split}
    J^\text{cr}_{ij}= & \,\omega_{\rm rec} \sum_{k=1}^M\Omega_{ik}\Omega_{jk} \cos{(\psi_{ik}-\psi_{jk})}
    \\& \times\left[\sum_{m\notin\mathcal{Q}}\frac{\xi_{im}\xi_{jm}}{\mu_k^2-\nu_m^2} + \sum_{\substack{q\in\mathcal{Q}\\q\neq k}}\frac{\xi_{iq}\xi_{jq}}{\mu_k^2-\mu_q^2}\frac{\mu_q}{\nu_q}\right].
\end{split}
\label{eq_Jcr}
\end{equation}
Here, $q\in\mathcal{Q}\equiv\{1,\ldots,M\}$ labels a motional mode assigned to a bichromatic drive, and $q\neq k$ selects an assigned mode other than the one addressed by drive $k$. Further details on the derivation of Eqs.~\eqref{eq_spin_phonon} and \eqref{eq_Jcr} can be found in the Supplemental Material. We note that we choose $\omega_k\equiv-\delta_k>0$ and $\psi_{ik}=0$ in the main text.

\subsection{Equivalence of  $\sigma_i^z$ and $n_i$ in the Hamiltonian}
\label{sec:app-spin_number_equivalence}

In this section, we show that a Hamiltonian in which the phonons couple to $\sigma^z_i$ rather than to the excitation number $n_i$,
\begin{align}
    \tilde H =& \sum_i \frac{\tilde\varepsilon_i}{2}\,\sigma^z_i
    +\sum_{i<j} J_{ij}\left(\sigma_i^+\sigma_j^- + \sigma^-_i\sigma^+_j\right) + \sum_{\lambda}\omega_{\lambda}a^\dagger_{\lambda}a_{\lambda}\nonumber\\
    &\quad + \sum_{i\lambda}\frac{g_{i\lambda}}{2}\,\sigma^z_i \left(a^\dagger_{\lambda}+a_{\lambda}\right),\label{eq:app_H_sigmaz}
\end{align}
is equivalent to Eq.~\eqref{eq_tot_H_ions}, up to a unitary displacement, a redefinition of the on-site potentials, and an additive constant. By identifying $n_i=\sigma^+_i\sigma^-_i=\left(1+\sigma^z_i\right)/2$, the spin--phonon coupling can be rewritten as
\begin{align}
    \sum_{i\lambda}\frac{g_{i\lambda}}{2}\sigma_i^z\left(a^\dagger_{\lambda}+a_{\lambda}\right)
    &=\sum_{i\lambda}g_{i\lambda} n_i\left(a^\dagger_{\lambda}+a_{\lambda}\right)\nonumber\\
    &\quad\quad-\frac{1}{2}\sum_{\lambda}G_{\lambda}\left(a^\dagger_{\lambda}+a_{\lambda}\right),
    \label{eq:app_sigmaz_to_n}
\end{align}
where $G_{\lambda}\equiv\sum_i g_{i\lambda}$. Thus, Eq.~\eqref{eq:app_H_sigmaz} becomes
\begin{align}
    \widetilde H
    =&\,\widetilde H_{\rm sp}+\sum_{\lambda}\omega_{\lambda}a^\dagger_{\lambda}a_{\lambda}+\sum_{i\lambda}g_{i\lambda}n_i\left(a^\dagger_{\lambda}+a_{\lambda}\right)\nonumber\\
    &-\frac{1}{2}\sum_{\lambda}G_{\lambda}\left(a^\dagger_{\lambda}+a_{\lambda}\right),
    \label{eq:app_Hprime}
\end{align}
with
\begin{equation}
    \widetilde H_{\rm sp}
    =
    \sum_i \frac{\widetilde\varepsilon_i}{2}\sigma_i^z
    +
    \sum_{i<j}J_{ij}
    \left(
        \sigma_i^+\sigma_j^-
        +\sigma_i^-\sigma_j^+
    \right).
\end{equation}
The final term in Eq.~\eqref{eq:app_Hprime} is independent of the spins and shifts the equilibrium position of each phonon mode. It can be removed by the displacement transformation
\begin{equation}
    U = \exp\left[\sum_{\lambda}\frac{G_{\lambda}}{2\omega_{\lambda}}
    \left(a^\dagger_{\lambda}-a_{\lambda}\right)\right],
    \label{eq:app_U_explicit}
\end{equation}
which acts on the ladder operators of each mode as a constant shift,
\begin{equation}
    U^\dagger a_{\lambda} U = a_{\lambda}+\frac{G_{\lambda}}{2\omega_{\lambda}},
    \qquad U^\dagger a^\dagger_{\lambda} U = a^\dagger_{\lambda}+\frac{G_{\lambda}}{2\omega_{\lambda}}.
    \label{eq:app_shift_explicit}
\end{equation}

Applying this transformation to Eq.~\eqref{eq:app_Hprime}, the
spin-independent terms linear in the phonon operators cancel exactly,
and one obtains
\begin{align}
    U^\dagger \widetilde H U=&\,\widetilde H_{\rm sp} +H_{\rm ph}+H_{\rm sp-ph}+\sum_i \Delta_i n_i-\sum_{\lambda}\frac{G_{\lambda}^2}{4\omega_{\lambda}},
    \label{eq:app_displaced_H}
\end{align}
where $\Delta_i\equiv\sum_{\lambda}\left(g_{i\lambda}G_{\lambda}/\omega_{\lambda}\right)=\sum_{j\lambda}\left(g_{i\lambda}g_{j\lambda}/\omega_{\lambda}\right)$. By using $n_i=(1+\sigma_i^z)/2$ again, the additional diagonal term can be written as
\begin{equation}
    \sum_i\Delta_i n_i=\sum_i\frac{\Delta_i}{2}\sigma_i^z+\frac{1}{2}\sum_i\Delta_i.
    \label{eq:app_Delta_n}
\end{equation}
The first contribution is absorbed into the on-site potentials by defining
\begin{equation}
    \varepsilon_i=\widetilde\varepsilon_i+\Delta_i=\widetilde\varepsilon_i +\sum_{j\lambda}\frac{g_{i\lambda}g_{j\lambda}}{\omega_{\lambda}}.
    \label{eq:app_epsilon_shift}
\end{equation}
Since $\sum_i\Delta_i=\sum_{\lambda}G_{\lambda}^2/\omega_{\lambda}$, the remaining constant is
\begin{equation}
    E_0=\sum_{\lambda}\frac{G_{\lambda}^2}{4\omega_{\lambda}}.
    \label{eq:app_constant_shift}
\end{equation}
Therefore, $U^\dagger \widetilde H U=H+E_0$, where $H$ is precisely the Hamiltonian of Eq.~\eqref{eq_tot_H_ions} with the renormalized on-site potentials in Eq.~\eqref{eq:app_epsilon_shift}. Equivalently, their time-evolution operators satisfy $e^{-iHt}=e^{iE_0t}\,U^\dagger e^{-i\widetilde Ht}U$. The phase $e^{iE_0t}$ is global and therefore does not affect physical observables. Hence, the $\sigma_i^z$-coupling and $n_i$-coupling formulations describe the same dynamics after the corresponding displacement and on-site-energy redefinition. We note that in the experimental implementation, $\varepsilon$ is realized using carrier Rabi drives (see Supplemental Material), which can exceed 100's kHz and can provide the correction required in Eq.~\eqref{eq:app_epsilon_shift}. For insufficient laser power, the efficient route is to directly null the spin-independent displacement in $H$ as discussed in Sec.~\ref{sec:Engineering_electron-phonon_models}.

\section{Lang--Firsov transformation}
\label{sec:app-Lang-firsov}

Starting from Eq.~\eqref{eq_tot_H_ions}, we apply the Lang--Firsov unitary $U_{\rm LF}=e^{S}$ of Eq.~\eqref{eq:langfirsov}, with generator
\begin{equation}
    S = \sum_i n_i B_i,
    \qquad
    B_i \equiv \sum_{\lambda}\frac{g_{i\lambda}}{\omega_{\lambda}}
    \left(a_{\lambda}-a^\dag_{\lambda}\right).
    \label{eq:appLF_S}
\end{equation}
The transformed operators are
\begin{eqnarray}
    U^\dag_{\rm LF}\,a_{\lambda}\,U_{\rm LF} &=& a_{\lambda}-\sum_i\frac{g_{i\lambda}}{\omega_\lambda}n_i,\\ 
    U^\dag_{\rm LF}\,\sigma_i^-\,U_{\rm LF} &=& \sigma_i^-\,e^{B_i},\\
    U_{\rm LF}^\dag\,\sigma^z_i\,U_{\rm LF} &=& \sigma^z_i.
    \label{eq:appLF_transform}
\end{eqnarray}
By substituting these relations into Eq.~\eqref{eq_tot_H_ions}, the linear spin--phonon coupling cancels exactly. In the single-excitation sector, $n_i^2=n_i$ and $n_i n_j=0$ for $i\neq j$, so the transformed Hamiltonian becomes
\begin{equation}
    \begin{split}
    H_{\rm LF} &= \sum_i\frac{\varepsilon_i}{2}\sigma^z_i 
    + \sum_{i<j}J_{ij}\left(\sigma_i^+\sigma_j^-\,e^{B_j-B_i}+{\rm h.c.}\right) \\
    &+ \sum_\lambda\omega_\lambda a^\dagger_\lambda a_\lambda
    - \sum_{i\lambda}\frac{g_{i\lambda}^2}{\omega_\lambda}\,n_i .
    \end{split}
    \label{eq:appLF_HLF}
\end{equation}
The residual coupling between the spin and phonon sectors is contained in the phonon-dressed hopping operator $e^{B_j-B_i}$. Its expectation value in the phonon vacuum is the Franck-Condon overlap
\begin{equation}
    \bra{0_{\rm ph}}e^{B_j-B_i}\ket{0_{\rm ph}}
    = \exp\left[-\frac{1}{2}\sum_\lambda
    \left|\frac{g_{i\lambda}-g_{j\lambda}}{\omega_\lambda}\right|^2\right].
    \label{eq:appLF_franckcondon}
\end{equation}
Projecting $H_{\rm LF}$ onto the zero-phonon sector therefore amounts to replacing $e^{B_j-B_i}$ by its phonon-vacuum expectation value, $\bra{0_{\rm ph}}e^{B_j-B_i}\ket{0_{\rm ph}}$. Since $\bra{0_{\rm ph}}a_\lambda^\dagger a_\lambda\ket{0_{\rm ph}}=0$, this gives the effective spin Hamiltonian in Eqs.~\eqref{eq:eff_hamiltonian}--\eqref{eq:eff_hopping}. The assumption of phonon vacuum is valid in the limit of $ J^{\rm eff}_{ij} \ll \omega_\lambda $. Since $J^{\rm eff}_{ij}$ is exponentially suppressed by Eq.~\eqref{eq:appLF_franckcondon}, this approximation becomes valid at strong coupling. In this regime, the spectrum of $H_{\rm LF}$ approximately splits into definite phonon-number excitations within the LF frame: the zero-phonon manifold gives the polaron band of $H^{\rm eff}$, while one-phonon states cost an extra $\sim\omega_\lambda$ and hybridize only weakly at order $J^{\rm eff}_{ij}\ll\omega_\lambda$. The first excited states of $H$ thus lie approximately $\omega_\lambda$ above the polaron ground state in this regime. 

\section{Open boundary conditions}
\label{sec:app-obc}

Consider a chain of $N$ sites with open boundary conditions and uniform nearest-neighbor
hopping,
\begin{align}
  H_{nn} &= \sum_{i<j} J_{ij} (\sigma_i^+ \sigma_j^- + \sigma^-_i\sigma^+_j)\\
  &= J\sum_{j=1}^{N-1}(\sigma_j^+ \sigma_{j+1}^- + \sigma^-_j\sigma^+_{j+1}).
\end{align}
Since the open chain is not translationally invariant, plane waves $e^{ikja}$ are not eigenstates of $H_{nn}$. In the single-excitation sector, the hopping Hamiltonian is instead diagonalized by standing-wave eigenmodes $\sin(k_n j a)$, with
\begin{equation}
  k_n = \frac{\pi n}{(N+1) a}, \qquad n = 1, \dots, N.
  \label{eq:kn}
\end{equation}
Accordingly, we define the momentum spin operators 
\begin{equation}
  \tilde{\sigma}^\pm_{k_n}
  = \sqrt{\frac{2}{N+1}} \sum_{j=1}^{N} \sin(k_n j a)\, \sigma_j^\pm.
  \label{eq:sine-transform}
\end{equation}
Inserting the operator definition into $H_{nn}$ and using the addition theorem
$\sin\!\big[k_n (j\pm1)a\big] = \sin(k_n j a)\cos(k_n a) \pm \cos(k_n j a)\sin(k_n a)$
together with the vanishing of the mode functions at $j = 0$ and $j=N+1$, we obtain 
\begin{equation}
  H_{nn} = \sum_{n=1}^{N} \varepsilon(k_n)\,
      \tilde{\sigma}^+_{k_n} \tilde{\sigma}^-_{k_n},
  \quad
  \varepsilon(k_n) = 2J \cos(k_n a) .
  \label{eq:dispersion}
\end{equation}
The dispersion in Eq.~\eqref{eq:dispersion} is identical to the periodic-chain result, and the open boundaries change only which values of $k$ are allowed, not the functional form of $\varepsilon(k)$. 

\clearpage                 
\onecolumngrid             
\setcounter{page}{1}       

\begingroup\centering{\large\bfseries Supplemental Material for Programmable Polaron Models and Tomography in Trapped-Ion Quantum Simulators}\par
\endgroup
\vskip -0pt   

\setcounter{equation}{0}
\setcounter{figure}{0}
\setcounter{table}{0}
\setcounter{section}{0}
\renewcommand{\theequation}{S\arabic{equation}}
\renewcommand{\thefigure}{S\arabic{figure}}
\renewcommand{\thetable}{S\arabic{table}}
\renewcommand{\theHequation}{S\arabic{equation}}
\renewcommand{\theHfigure}{S\arabic{figure}}
\renewcommand{\theHtable}{S\arabic{table}}

\setcounter{section}{0}
\setcounter{subsection}{0}
\setcounter{secnumdepth}{3}                      
\renewcommand{\thesection}{S\arabic{section}}
\renewcommand{\thesubsection}{\thesection.\arabic{subsection}}
\renewcommand{\thesubsubsection}{\thesubsection.\arabic{subsubsection}}

\newcommand{\suppsection}[1]{%
  \refstepcounter{section}%
  \section*{\thesection. #1}%
}

\counterwithout{equation}{section}     
\setcounter{equation}{0}               
\renewcommand{\theequation}{S\arabic{equation}}
\renewcommand{\theHequation}{S\arabic{equation}}

\suppsection{Local phonon mapping}

In this section, we discuss the general \emph{passive} Bogoliubov transformation of the normal phonon mode basis to the local phonon mode basis as referred to in Sec. II of the main text. The criteria required for the new local phonon operators for the models considered in this work are the following: {\bf\emph{(i)}} orthogonality of annihilation and creation operators via commutation relations, {\bf\emph{(ii)}} ability to realize dispersionless phonons, and {\bf\emph{(iii)}} linear transformation between linear bosonic observables and spin-phonon couplings. In other words, local mode ladder operators $a_m$ and $a^{\dagger}_{m'}$ need to satisfy
\begin{equation}
    \label{eq:localmoderequirements}
    [a_m,a^\dagger_{m'}]=\delta_{m,{m'}},
    \quad
    H_{\text{ph}}=\sum_\lambda \omega_{\lambda} a^{\dagger}_\lambda a_\lambda= \sum_m \omega_m a^{\dagger}_m a_m,
    \quad
    H_{\rm sp-ph}=\sum_{i\lambda} g_{i\lambda} n_i x_\lambda = 
    \sum_{im} \tilde{g}_{im} n_i x_m,
\end{equation}
where $\omega_\lambda=\omega_m=\omega_0$ is the common detuning set in the interaction frame, and $x_{\lambda}\equiv (a^{\dagger}_\lambda+a_\lambda)$ is the normal mode position operator.

A natural transformation between the normal-mode and local-mode descriptions of the ion lattice is defined by the orthogonal eigenvector matrix of the radial trapping potential, $\xi_{m\lambda}$, where $m=1,\ldots,N$ labels the ion sites, and $\lambda=1,\ldots,M$ labels the addressed normal modes. Although one may write $x_m=\sum_{\lambda=1}^{M}\xi_{m\lambda}x_\lambda$, the three criteria above can be satisfied simultaneously only when the complete set of normal modes is included, i.e., when $M=N$. For $M<N$ normal mode phonons addressed, the lesser number of non-zero columns in $\xi_{m\lambda}$ breaks operator orthogonality, prevents the realization of a dispersionless phonon spectrum, and inhibits exact solutions for $\tilde{g}_{im}$.

There are two approaches to circumvent these limitations. First, we can use the symmetric Lo\"wdin orthogonalization \cite{lowdin1950orthogonalization} to construct a new basis of quasi-local Wannier-like phonon modes from the normal mode basis that fulfills the three criteria above, even when $M\leq N$.  Alternatively, one may retain the full $N\times N$ transformation matrix $\xi_{m\lambda}$, thereby preserving operator orthogonality and a dispersionless phonon spectrum, while allowing approximate solutions for $\tilde{g}_{im}$ (see Sec.~\ref{sec:RLPmapping}).

\subsection{L\"owdin orthogonalization}

We consider the Lo\"wdin orthogonalization of the rectangular coupling matrix $g_{i\lambda}$ of size $N_{\rm s} \times M$, where $N_{\rm s}$ is the number of ions addressed, and $N_{\rm s}\leq M$. The coupling matrix can then be decomposed as
\begin{equation}
\label{eq:Lowdinorthogonalization}
    g=LV,
    \qquad 
    L=gV^{T},
\end{equation}
where $L$ can be any target matrix of size $N_s\times N_s$, and $V$ is the transformation matrix (also called gauge matrix) of size $N_{\rm s}\times M$ with orthogonal rows, i.e. $VV^T=\mathbb{I}_{N_{\rm s}}$ (right orthogonal). This implies an exact solution for $L$ exists for a unique combination of $V$ and $g$. In the special case of symmetric L\"owdin orthogonalization, $L$ is symmetric and positive-semi-definite, and follows $L=(gg^T)^{1/2}$. Consequently, the orthogonality of $V$ follows, $VV^T=(gg^T)^{-1/2}g (g^T (gg^T)^{-1/2}) =\mathbb{I}_{N_{\rm s}}$. The properties of the general L\"owdin orthgonalization allow us to define a set of Wannier-like local phonon operators (WLP) in terms of the normal mode operators, such as annihilation operators $a_m=\sum^M_{\lambda=1} V_{m\lambda} a_{\lambda}$. These operators satisfy the following properties:
\begin{enumerate}[label={(\roman*)}]
    \item \emph{Orthogonality}: Starting from the commutation relations of the Wannier-like phonon operators, we get
    \begin{align}
    \left[a_m, a^\dagger_{m{'}}\right]=& \,\left[\sum^M_{\lambda=1} V_{m\lambda} a_{\lambda},\, \sum^M_{\lambda{'}=1} V_{\lambda{'}m{'}} a^{\dagger}_{\lambda{'}}\right]
    \nonumber \\= &\,\sum^M_{\lambda=1} \sum^M_{\lambda{'}=1}V_{m\lambda}V_{\lambda{'}m{'}}\left[a_{\lambda},\, a^{\dagger}_{\lambda{'}}\right]
     \nonumber \\=& \, \sum^M_{\lambda=1} \sum^M_{\lambda{'}=1}V_{m\lambda}V_{\lambda{'}m{'}}\,\delta_{\lambda\lambda{'}}
     \nonumber \\=& \, \sum^M_{\lambda=1} V_{m\lambda}V_{\lambda m{'}}= VV^T
     \nonumber \\=&\, \delta_{m m{'}}.
    \label{eq:wannierorthogonality}
    \end{align}
    
    \item \emph{Dispersionless phonon spectrum}: $V$ consists of $N_{\rm s}$ orthogonal vectors in the subspace of dimension $M$, which decomposes $a_m$ in terms of $a_\lambda$. However, there is an orthogonal subspace of rank $M-N_{\rm s}$ and a corresponding matrix $W$ of size $(M-N_{\rm s}) \times M$ that completes the entire vector space of dimension $M$ and follows
    \begin{equation}
    \label{eq:orthogonalsubspaceV}
    O=\begin{pmatrix}
        V_{N_{\rm s}\times M} \\
        W_{(M-N_{\rm s})\times M}
        \end{pmatrix},
    \qquad
    VV^T=\mathbb{I}_{N_{\rm s}},
    \qquad 
    WW^T=\mathbb{I}_{M-N_{\rm s}},
    \qquad
    VW^T=\boldsymbol{0},
    \end{equation}
    where $O$ is the corresponding orthogonal matrix of size $M\times M$. The complete set of orthogonal vectors in the $M$-dimensional space allows
    \begin{equation}
    \label{eq:orthogonaltransformWannier}
    \vec{d}=O\vec{a_\lambda},
    \qquad
    \vec{a_\lambda}=O^T\vec{d},
    \qquad
    a_\lambda=\sum^{N_{\rm s}}_{m=1} V_{\lambda m} a_{m} + \sum^{M-N_{\rm s}}_{\alpha=1} W_{\lambda\alpha} a_{\alpha} , 
    \end{equation}
    where $\vec{d}$ is the vector consisting of all the annihilation operators of Wannier-like phonons. The operators $a_\alpha$ constitute the spectator set of Wannier-like phonons in the $M$-dimensional space orthogonal to $a_m$, and these $a_\alpha$ phonon modes are unaddressed. Substituting Eq.~\eqref{eq:orthogonaltransformWannier} into $H_{\rm ph}$ of Eq.~\eqref{eq:localmoderequirements} for $\omega_{\lambda}=\omega_0$, we get
    \begin{align}
    H_{\rm ph}=&\,\omega_0\sum^M_{\lambda=1} a^\dagger_\lambda a_\lambda
    \nonumber \\=&\,\omega_0 \sum^M_{\lambda=1}\left(\sum^{N_{\rm s}}_{m=1} V_{m \lambda} a^\dagger_{m} + \sum^{M-N_{\rm s}}_{\alpha=1} W_{\alpha \lambda} a^\dagger_{\alpha}\right)\left(\sum^{N_{\rm s}}_{m{'}=1} V_{\lambda m{'}} a_{m {'}} + \sum^{M-N_{\rm s}}_{\alpha{'}=1} W_{\lambda\alpha{'}} a_{\alpha{'}}\right)
    \nonumber\\=&\, \omega_0 \sum^{M}_{\lambda=1} \left(  \sum^{N_{\rm s}}_{m=1} \sum^{N_{\rm s}}_{m{'}=1} V_{m \lambda}  V_{ \lambda m {'}} a^\dagger_{m}  a_{m{'}}
    + \sum^{N_{\rm s}}_{m=1}\sum^{M-N_{\rm s}}_{\alpha{'}=1} V_{m \lambda}  W_{ \lambda \alpha {'}} a^\dagger_{m}  a_{\alpha{'}}
    \right.\nonumber \\&+\left.\sum^{M-N_{\rm s}}_{\alpha=1}\sum^{N_{\rm s}}_{m{'}=1} W_{\alpha \lambda}  V_{ \lambda m {'}} a^\dagger_{\alpha}  a_{m{'}}
    + \sum^{M-N_{\rm s}}_{\alpha=1}\sum^{M-N_{\rm s}}_{\alpha{'}=1} W_{\alpha \lambda}  W_{ \lambda \alpha {'}} a^\dagger_{\alpha}  a_{\alpha{'}}
    \right)
    \nonumber\\=&\, \omega_0\left( \sum^{N_{\rm s}}_{m=1} a^{\dagger}_m a_m 
    + \sum^{M-N_{\rm s}}_{\alpha=1} a^{\dagger}_{\alpha} a_{\alpha}  \right).
    \label{eq:wannierdispersionless}
    \end{align}
    Restricting to the addressed Wannier-like phonon subspace therefore gives the desired dispersionless phonon spectrum, $H_{\rm ph}=\omega_0\sum^{N_{\rm s}}_{m=1}a^\dagger_m a_m$.    
    \vspace{-0.5cm}
    \item \emph{Linear transformation of spin--phonon coupling}: We can re-express $H_{\rm sp-ph}$ in terms of the Wannier-like phonon operators as
    \begin{align}
    H_{\rm sp-ph}=& \sum^{N_{\rm s}}_{i=1}\sum^{M}_{\lambda=1} g_{i\lambda} n_i \left(a^\dagger_\lambda + a_\lambda\right)
    \nonumber \\= &\, \sum^{N_{\rm s}}_{i=1}\sum^{M}_{\lambda=1} g_{i\lambda} n_i \left(\sum^{N_{\rm s}}_{m=1} V_{\lambda m} (a^\dagger_{m} + a_{m})  
    +\sum^{M-N_{\rm s}}_{\alpha=1} W_{\lambda \alpha} (a^\dagger_{\alpha} + a_{\alpha})\right).
    \end{align}
     By ignoring the spectator modes ($a_\alpha$ and $a^\dagger_\alpha$), the expression simplifies to
    \begin{align}
    H_{\rm sp-ph}= &\, \sum^{N_{\rm s}}_{i=1}\sum^{M}_{\lambda=1} g_{i\lambda} n_i \left(\sum^{N_{\rm s}}_{m=1} V_{\lambda m} \left(a^\dagger_{m} + a_{m}\right)\right)
    \nonumber \\= &\,\sum^{N_{\rm s}}_{i=1} \sum^{N_{\rm s}}_{m=1} \left( \sum^{M}_{\lambda=1}  g_{i\lambda}V_{\lambda m} \right) n_i  \left(a^\dagger_{m} + a_{m}\right)
    \nonumber \\= &\, \sum^{N_{\rm s}}_{i=1}\sum^{N_{\rm s}}_{m=1} L_{im}n_i  \left(a^\dagger_{m} + a_{m}\right)
    \nonumber \\\equiv &\, \sum^{N_{\rm s}}_{i=1}\sum^{N_{\rm s}}_{m=1} \tilde{g}_{im} n_i \left(a^\dagger_{m} + a_{m}\right).
    \label{eq:lineargtransform}
    \end{align}
    For a known $V$, the equivalence of $ \tilde{g}_{im}=L_{im}=\sum^M_{\lambda=1} g_{i\lambda}V_{\lambda m}$ implies that, for a target spin-local phonon coupling $\tilde{g}_{im}$, there is an exact unique solution to the required ${g}_{i\lambda}$ to be programmed in the trapped-ion simulator.   
\end{enumerate}

\subsection{Extracting optimal $g_{i\lambda}$ for target $\tilde{g}_{im}$}

Given that the target interaction is $L=\tilde{g}_{im}$, to determine $g_{i\lambda}$, we first need to construct $V$ belonging to the same subspace as $g_{i\lambda}$. We obtain an orthogonal basis in the $N_{\rm s}$ subspace that closely resembles $\xi_{m\lambda}$ by first extracting the singular value decomposition (SVD) of $\xi^{\rm S}_{m\lambda}$ of size $N_{\rm s} \times M$, as follows
\begin{equation}
\xi^{\rm S}= u w v^T,
\qquad 
L'\equiv(\xi^{\rm S} (\xi^{\rm S})^T)^{1/2},
\label{eq:svdchi}
\end{equation}
where $u$ is the orthogonal matrix of size $N_{\rm s}\times N_{\rm s}$, $w$ is the singular value matrix of $\xi^{\rm S}$, and $v$ is the rectangular matrix of size $M\times N_{\rm s}$ whose columns are orthonormal to one another. $L'$ is the newly defined symmetric L\"owdin matrix such that $\xi^{\rm S}=L' V$. Since $(L')^2= uw^2u^T$, and $L'= uwu^T$,  
$V$ is obtained from
\begin{align}
V=&\, (L')^{-1} \xi^{\rm S}
\nonumber \\=&\, (uw^{-1}u^T) u w v^T
\nonumber \\=&\, uv^T.
\label{eq:derivingV}
\end{align}
Hence, substituting the newly found $V$ into $g_{i\lambda}=\sum^{N_{\rm s}}_{m=1} \tilde{g}_{im} V_{m\lambda}$ provides an exact solution of $g_{i\lambda}$ for given $\tilde{g}_{im}$ and $V$. Since $g_{i\lambda}=\eta_\lambda\xi_{i\lambda}\Omega_{i\lambda}$, the magnitude of the Rabi frequency $\Omega_{i\lambda}$ can be tuned to satisfy the solution of $g_{i \lambda}$, where the sign can be optionally tuned with the spin phase $\phi_{i\lambda}$ per ion. To remain within the power budget allocated per laser tone, $\Omega_{i\lambda}$ is renormalized by $\Omega^{\rm max}$.

\subsection{Observables}
Linear observables of L\"owdin-orthogonalized local phonons (e.g., $\braket{x_m}$ and $\braket{\sigma^\phi_i p_m}$) can be retrieved by transforming the measured linear normal mode observables using $V$, as described in the main text. However, evaluating higher-order observables, such as $a^\dagger_m a_m$, requires measuring correlations between multiple normal-mode phonon operators, which is experimentally more challenging. In contrast, due to the direct mapping between synthetic local phonons and normal modes in the SLP mapping, all bosonic and spin--boson observables of the local modes can be directly measured through operations on the corresponding normal modes.

\subsection{Phonon locality}
The centroid and spatial spread of the Wannier-like phonons are determined by $V$, together with the choice of $N_{\rm s}$ ions and $M$ addressed modes. As $M$ approaches $N$, the centroids of the Wannier-like phonons converge to the ion-lattice positions $i$ with point-like spatial spread. The spatial overlap of the Wannier-like phonons and the ion-site local phonons can be quantified with the spatial overlap matrix $\Pi$ of size $N_{\rm s}\times N$ defined as
\begin{equation}
\Pi=V\xi^T,
\label{eq:Wannierspatialoverlap}
\end{equation}
where $\xi$ denotes the $N\times M$ restriction of the normal-mode matrix to
the $M$ driven columns, retaining all $N$ ion rows.
Each row of $\Pi$ is automatically a valid probability distribution over the
site index, with no separate normalization step required. Using
$\xi^{T}\xi=\mathbb{I}_M$ (any subset of columns of a full orthogonal matrix
remains orthonormal) and $VV^{T}=\mathbb{I}_{N_{\rm s}}$, we obtain
\begin{equation}
\Pi\,\Pi^{T}=V\big(\xi^{T}\xi\big)V^{T}=VV^{T}=\mathbb{I}_{N_{\rm s}}.
\label{eq:Pi_orthonormal}
\end{equation}
In particular, $\sum_{i=1}^{N}(\Pi_{mi})^{2}=1$ for every $m$. Hence,
$p_m(i)\equiv(\Pi_{mi})^{2}$ defines a genuine probability distribution over
sites, and the rows for different Wannier-like modes are mutually orthogonal.

The Wannier centroid (center of mass on the ion lattice) and its spatial
spread follow directly from the first and second moments of $p_m(i)$,
\begin{equation}
\label{eq:centroid}
\bar i_m\equiv\langle {i}\rangle_m=\sum_{i=1}^{N}i\,(\Pi_{mi})^{2},
\end{equation}
\begin{equation}
\label{eq:spread}
\sigma_m\equiv\sqrt{\langle i^{2}\rangle_m-\langle i\rangle_m^{2}}
=\sqrt{\sum_{i=1}^{N}i^{2}(\Pi_{mi})^{2}-(\bar i_m)^{2}},
\end{equation}
with distance measured in ion-site units. The offset of the centroid from the nominal site $i_m$, $\Delta_m\equiv\bar{i}_m-i_m$, and the spatial spread $\sigma_m$ quantify, respectively, how far and how broadly a given Wannier phonon deviates from its nominal ion site. As $M$ addressed modes approaches $N$, $p_m(i_m)\rightarrow1$, $\bar i_m\rightarrow i_m$, and $\sigma_m\rightarrow0$. 

\suppsection{Real-Space local phonon mapping
\label{sec:RLPmapping}
}

In this section, we discuss the RLP mapping of the normal phonon mode basis to the drive-induced real-space local phonon mode basis as referred to in Sec. II of the main text.
Consider the local phonon modes,
\begin{equation}
\label{eq:normal_mode_basis}
    x_{m}=\sum_{\lambda=1}^{N}\xi_{m \lambda}x_{\lambda},
\end{equation}
where the dimensionless normal mode position operator is $x_{\lambda}\equiv a^{\dagger}_\lambda+a_\lambda$. The dimensionless local mode position operator are also defined as $x_{m}\equiv a^{\dag}_{m}+ a_{m}$, and the corresponding local mode momentum operator is $p_{m}\equiv i(a^{\dagger}_{m}- a_{m})$, similar to the normal mode basis. Due to the orthogonality of $\xi_{m\lambda}$, the operators in the local mode basis inherit the canonical commutation relations from the normal mode basis, $[x_m,p_{m'}]=2i\delta_{m,m'}$, and $[a_m,a^\dagger_{m'}]=\delta_{m,m'}$, which generalize the transformation in Eq.~\eqref{eq:normal_mode_basis} to $\square_m=\sum_{\lambda=1}^{N}\xi_{m \lambda}\square_{\lambda}$, where $\square\in\{x,p,a, a^\dagger,\,\text{other linear operators}\}$.

We can invert the orthogonal normal mode eigenvector matrix $\xi_{m\lambda}$ to represent the normal modes, $\{x_{\lambda}\}$, in terms of the local phonon modes, $\{x_{m}\}$, as
\begin{equation}
\label{eq:local_mode_basis}
    x_{\lambda}=\sum_{m=1}^{N} \xi_{\lambda m} x_m,
\end{equation}
where $\xi_{\lambda m}\equiv(\xi^{T})_{\lambda m}$, and $\sum_m\xi_{\lambda m} \xi_{m \lambda'}=\delta_{\lambda \lambda'}$.
Rewriting $H_{\rm sp-ph}$ in Eq.~(1) of the main text, we get
\begin{align}
    \label{eq:rewritten_H_sp-ph}
    H_{\rm sp-ph}&=
    \sum^{N}_{i=1} \sum^{N}_{\lambda=1} \eta_{\lambda}\Omega_{i\lambda} \xi_{i\lambda}  (n_{i}x_{\lambda})\nonumber\\
    &\equiv
    \sum^{N}_{i=1} \sum^{N}_{\lambda=1} g_{i\lambda}  (n_{i}x_{\lambda}).
\end{align}
Here, $\Omega_{i\lambda}$ is the carrier Rabi frequency corresponding to the off-resonant bichromatic tone applied on ion \emph{i} for mode $\lambda$ and is independently tunable with individual addressing. Further, $\Omega_{i\lambda}\neq0$ for only $M\leq N$ normal modes, and thus only $M$ columns of $g_{i\lambda}$ are non-zero. Substituting Eq.~\eqref{eq:local_mode_basis} into Eq.~\eqref{eq:rewritten_H_sp-ph} yields
\begin{equation}
    \label{H_el-ph_local_modes}
H_{\rm sp-ph}=\sum^{N}_{i=1} \sum^{N}_{m=1}  \left(\sum^{N}_{\lambda=1}  \xi_{i \lambda} \xi_{\lambda m} \eta_{\lambda}\Omega_{i\lambda} \right) \left(n_{i}x_{m}\right).
\end{equation}
Defining $\tilde{g}_{im}=\sum^{N}_{\lambda=1} 
\eta_{\lambda} \Omega_{i\lambda}  \xi_{i\lambda} \xi_{\lambda m}= \sum^{N}_{\lambda=1} 
g_{i\lambda}\xi_{\lambda m}$ results in
\begin{equation}
\label{eq:localphononcoupling}
H_{\rm sp-ph}=\sum^{N}_{i=1} \sum^{N}_{m=1} \tilde{g}_{im} n_{i}x_{m}.  
\end{equation}

If we take $\Omega_{i\lambda}= \Omega_0/\eta_{\lambda}$ across all the ions, then, for $M=N$, $\tilde{g}_{im}$  is proportional to identity and trivially describes an ideal Holstein model. This suggests that driving all the normal modes near resonance with detuning $\omega_\lambda=\omega_0$ couples each local phonon locally to each ion at the corresponding site in real space.
The variation of $\tilde{g}_{im}$  with respect to different modes is depicted in Fig.~2 of the main text.

\subsection{Optimizing $\Omega_{i\lambda}$ to achieve a target $\tilde{g}_{im}$ profile}

Since each Rabi frequency applied per ion per normal mode, $\Omega_{i\lambda}$, is tunable, we can also tailor the local phonon-spin coupling matrix, $\tilde{g}_{im}$, to an exponential or power-law profile for a given number of modes. Although using $M=N$ normal modes provides a direct mapping to the target $\tilde{g}^{\rm target}_{im}$, this implementation can be inhibiting due to the finite laser power budget. Hence, a smaller number of optimal modes can be considered to approximate the target profile. Exponential and power-law profiles more closely match the target when the characteristic interaction length extends beyond nearest neighbors, favoring longer-range interactions of the Fr\"ohlich type, which arise more naturally in trapped-ion systems.

\vspace{0.43cm}
Optimization procedure:
\begin{enumerate}
    \item Define the normalized $\tilde{g}^{\rm target}_{im}$ matrix of size $N_{\rm s} \times N_{\rm s}$, where max($\tilde{g}^{\rm target}_{im}$)=1.
    \item Define a function $\tilde{g}_{im}$ based on Eq.~\eqref{H_el-ph_local_modes} with variable $\Omega_{i\lambda}$ matrix of size $N_{\rm s}\times M$, consisting of non-zero columns.
    \item Create an outer loop that iterates over different combinations of $M$ modes and evaluates $\tilde{g}_{im}$ with variable $\Omega_{i \lambda}$.
    \item Minimize the cost function, which we choose to be the squared Frobenius norm  $(||\tilde{g}^{\rm target}_{im}-\tilde{g}_{im}||_2)^2$, in the inner loop. For each mode configuration, we obtain the $\Omega_{i\lambda}$ matrix, and hence $g_{i\lambda}$, resulting in  $\tilde{g}_{im}$ closest to $\tilde{g}^{\rm target}_{im}$. Additionally, we apply constraints to maximize the magnitude and uniformity of the diagonal elements of $\tilde{g}_{im}$ and ensure non-negativity in $\tilde{g}_{im}$ and $\Omega_{i\lambda}$ for the models considered here.
    \item Sorting over all these local optimal configurations, the mode configuration with the lowest cost is obtained, whose optimal $\Omega^{\rm opt}_{i\lambda}$ results in the spin--phonon coupling $g^{\rm opt}_{i\lambda}$ tailored for a $\tilde{g}^{\rm target}_{im}$ profile.
    \item Renormalize $\Omega^{\rm opt}_{i\lambda}$ by $\Omega^{\rm max}$, which is the maximum carrier Rabi frequency corresponding to each laser tone addressing the ion chain. This results in the realistic maximum $\tilde{g}^{\rm target}_{im}$, which is equal to or lower than what was initially set before the renormalization.
\end{enumerate}

\subsection{Phonon dispersion in interaction frame}

A consequence of the real-space Bogoliubov transformation in Eq.~\eqref{eq:normal_mode_basis} is the existence of phonon--phonon hopping terms arising in $H_{\rm ph}$ when $\omega_\lambda \neq \omega_0$. The phonon sector of the Hamiltonian in Eq.~(1) of the main text, $H_{\rm ph}=\sum^M_{\lambda=1} \omega_\lambda a_{\lambda}^{\dagger }a_{\lambda}$, transforms to
\begin{align}
\label{eq:ph_ph_hopping}
    H_{\rm ph}= &\sum_{\lambda\,m\,m'} \xi_{m \lambda} \omega_\lambda \xi_{\lambda m'}a^\dagger_{m} a_{m'} 
   \nonumber \\=&   \sum_{m\,m'} h_{mm'} a^\dagger_{m} a_{m'} \,,
\end{align}
where $h=\xi\text{diag}(\omega_\lambda)\xi^{T}$. In the case of $\omega_\lambda= \omega_0$ for dispersionless optical phonons, the phonon--phonon hopping terms are eliminated as $h$ is simply an identity matrix scaled by $\omega_0$. This restricts the independent tunability of the dispersion profile of the simulatable phonon spectrum. However, the synthetic-space mapping of normal mode to local mode phonons circumvents this restriction as it directly assigns $\tilde{g}_{im}=g_{i\lambda}$ without any transformation. This allows $\omega_\lambda$ and hence $\omega_m$ to be tuned independently for phonon dispersion.

\suppsection{Derivation of multiple simultaneous spin--phonon drives}

We consider a chain of $N$ ions driven by $M\leq N$ bichromatic Raman fields, with site-resolved control of their amplitudes and phases. The Raman wavevector difference $\Delta\vec{k}$ is taken along a radial direction orthogonal to the one used to generate the long-range spin-hopping interactions discussed in the main text. Throughout this section, we set $\hbar=1$.

Each bichromatic drive $k$ consists of two beatnotes symmetrically detuned from the qubit resonance, $\omega_{k,B}^{L}=\omega_0+\mu_k$ and $\omega_{k,R}^{L}=\omega_0-\mu_k$, with equal Rabi frequency $\Omega_{ik}$ on ion $i$ and independently controllable optical phases $\varphi_{ik}^{B}$ and $\varphi_{ik}^{R}$. The total Hamiltonian of the ion chain is ($\hbar=1$)
\begin{equation}
    H=\underbrace{\sum_{i=1}^{N}\frac{\omega_0}{2}\sigma_i^z}_{H_A}
    +\underbrace{\sum_{m=1}^{N}\nu_m a_m^\dagger a_m}_{H_m}   
    +\underbrace{\sum_{i=1}^{N}\sum_{k=1}^{M}\sum_{s=B,R}\frac{\Omega_{ik}}{2}(\sigma_i^++\sigma_i^-)\left[e^{i(\Delta k x_i-\omega_{k,s}^{L}t-\varphi_{ik}^{s})}+e^{-i(\Delta k x_i-\omega_{k,s}^{L}t-\varphi_{ik}^{s})}\right]}_{H_I}.
\label{eq:supp_lab_H}
\end{equation}
Here, $i$ labels the ions, $m$ labels the collective normal modes of the relevant radial branch, and $k$ labels the bichromatic drives. Neglecting micromotion, the displacement of ion $i$ along the direction of $\Delta\vec{k}$ gives $\Delta k x_i=\sum_{m=1}^{N}\eta_{im}(a_m+a_m^\dagger)$, where $\eta_{im}=\eta_m\xi_{im}$ and $\eta_m = \Delta k \sqrt{\frac{1}{2m_{\rm ion}\nu_m}}$ are the Lamb--Dicke parameters, $\xi_{im}$ is the orthogonal normal-mode matrix, and $m_{\rm ion}$ is the ion mass~\cite{leibfried2003quantum,Monroe2021}.

We choose the $M$ bichromatic drives to be near resonant with a subset of $M$ motional modes, $\mathcal Q\equiv\{1,\ldots,M\}$, and label the drives such that drive $k$ predominantly addresses mode $k$. We define the corresponding detuning $\delta_k\equiv\mu_k-\nu_k$. To work in a frame in which the desired spin--phonon interactions are stationary, we apply a unitary transformation with respect to
\begin{equation}
    U(t)=e^{-i(H_A+H_m+H_{\mathcal Q})t},
\end{equation}
where we have introduced $H_{\mathcal Q}\equiv\sum_{q\in\mathcal Q}\delta_q a_q^\dagger a_q$~\cite{Schneider_2012,so2025,So2026Thesis}. The Hamiltonian in this frame thus becomes
\begin{equation}
\begin{split}
    H^{(\rm res)}&=i\left(\frac{\partial U^\dagger}{\partial t}\right)U+U^\dagger H U\\
    &=U^\dagger H_I U-H_{\mathcal Q}\\
    &\equiv H_I^{(\rm res)}-H_{\mathcal Q},
\end{split}
\label{eq:supp_res_frame}
\end{equation}
The term $-H_{\mathcal Q}$ provides the effective phonon Hamiltonian $H_{\rm ph}$ in the main text. For $\omega_k\equiv-\delta_k$, we get
\begin{equation}
    -H_{\mathcal Q}=\sum_{k=1}^{M}\omega_k a_k^\dagger a_k.
\label{eq:supp_Hph}
\end{equation}
Here, we choose $\delta_k<0$, such that $\omega_k>0$. We emphasize that $\omega_k$ is an effective phonon frequency given by the laser detuning and is distinct from the physical normal-mode frequency $\nu_k$.

In this rotating frame, the motional displacement appearing in the light--matter interaction described by $H_I^{(\rm res)}$ evolves as
\begin{equation}
    \Delta k x_i(t)=\sum_{m\notin\mathcal Q}\eta_{im}\left(a_m e^{-i\nu_m t}+a_m^\dagger e^{i\nu_m t}\right)+\sum_{q\in\mathcal Q}\eta_{iq}\left(a_q e^{-i\mu_q t}+a_q^\dagger e^{i\mu_q t}\right).
\label{eq:supp_motion_res}
\end{equation}

We further apply the optical rotating-wave approximation and neglect terms oscillating at frequencies $\omega_{k,s}^{L}+\omega_0$. By working in the Lamb--Dicke regime, $\eta_{im}\sqrt{\left\langle(a_m+a_m^\dagger)^2\right\rangle}\ll1$, and assuming $|\mu_k|\gg|\Omega_{ik}|$, we expand the interaction to first order in the Lamb--Dicke parameters and obtain
\begin{equation}
    H_I^{(\rm res)}=\sum_{i=1}^{N}\sum_{k=1}^{M}\Omega_{ik}\cos(\mu_k t+\psi_{ik})\sigma_i^{\phi_{ik}}\left[\sum_{m\notin\mathcal Q}\eta_{im}\left(a_m e^{-i\nu_m t}+a_m^\dagger e^{i\nu_m t}\right)+\sum_{q\in\mathcal Q}\eta_{iq}\left(a_q e^{-i\mu_q t}+a_q^\dagger e^{i\mu_q t}\right)\right],
\label{eq:supp_bichromatic}
\end{equation}
where $\phi_{ik}\equiv(\varphi_{ik}^{B}+\varphi_{ik}^{R})/2-\pi/2$ and $\psi_{ik}\equiv(\varphi_{ik}^{B}-\varphi_{ik}^{R})/2$ are the spin and motional phases, respectively, and $\sigma_i^{\phi_{ik}}\equiv\sigma_i^x\cos\phi_{ik}+\sigma_i^y\sin\phi_{ik}$~\cite{leibfried2003quantum,Monroe2021}. We now separate the second sum in Eq.~\eqref{eq:supp_bichromatic} into the target contribution $q=k$ and the off-resonant contributions $q\neq k$. For the target mode,
\begin{equation}
\begin{split}
    \eta_{ik}\Omega_{ik}\cos(\mu_k t+\psi_{ik})\sigma_i^{\phi_{ik}}\left(a_k e^{-i\mu_k t}+a_k^\dagger e^{i\mu_k t}\right)
    =&\frac{\eta_{ik}\Omega_{ik}}{2}\sigma_i^{\phi_{ik}}\left(a_k e^{i\psi_{ik}}+a_k^\dagger e^{-i\psi_{ik}}\right)\\
    &+\frac{\eta_{ik}\Omega_{ik}}{2}\sigma_i^{\phi_{ik}}\left(a_k e^{-i(2\mu_k t+\psi_{ik})}+a_k^\dagger e^{i(2\mu_k t+\psi_{ik})}\right).
\end{split}
\label{eq:supp_target_mode}
\end{equation}
The second contribution oscillates at $\pm2\mu_k$ and is again neglected under the corresponding rotating-wave approximation. By defining $g_{ik}\equiv\eta_{ik}\Omega_{ik}$, we obtain the desired resonant interaction
\begin{equation}
    H_{\rm sp-ph}=\sum_{i=1}^{N}\sum_{k=1}^{M}\frac{g_{ik}}{2}\sigma_i^{\phi_{ik}}\left(a_k e^{i\psi_{ik}}+a_k^\dagger e^{-i\psi_{ik}}\right).
\label{eq:supp_Hsp}
\end{equation}
The remaining off-resonant terms are
\begin{equation}
\begin{split}
    H_{\rm cr}(t)=&\sum_{i=1}^{N}\sum_{k=1}^{M}\sum_{m\notin\mathcal Q}\eta_{im}\Omega_{ik}\cos(\mu_k t+\psi_{ik})\sigma_i^{\phi_{ik}}\left(a_m e^{-i\nu_m t}+a_m^\dagger e^{i\nu_m t}\right)\\
    &+\sum_{i=1}^{N}\sum_{k=1}^{M}\sum_{\substack{q\in\mathcal Q\\q\neq k}}\eta_{iq}\Omega_{ik}\cos(\mu_k t+\psi_{ik})\sigma_i^{\phi_{ik}}\left(a_q e^{-i\mu_q t}+a_q^\dagger e^{i\mu_q t}\right).
\end{split}
\label{eq:supp_Hcr_t}
\end{equation}
The first term describes the off-resonant action of each bichromatic drive on the spectator modes $m\notin\mathcal Q$, while the second describes its off-resonant action on addressed modes $q\neq k$ assigned to other bichromatic drives.

Together with the phonon Hamiltonian in Eq.~\eqref{eq:supp_Hph}, the full Hamiltonian in the resonant frame is therefore
\begin{equation}
    H^{(\rm res)}=\sum_{i=1}^{N}\sum_{k=1}^{M}\frac{g_{ik}}{2}\sigma_i^{\phi_{ik}}\left(a_k e^{i\psi_{ik}}+a_k^\dagger e^{-i\psi_{ik}}\right)+\sum_{k=1}^{M}\omega_k a_k^\dagger a_k+H_{\rm cr}(t).
\label{eq:supp_Hrot_full}
\end{equation}

We can further simplify the effective interaction generated by $H_{\rm cr}(t)$ by considering the spectrally resolved regime, where $|\mu_k-\nu_m|\gg|\eta_{im}\Omega_{ik}|,\,m\notin\mathcal Q$ and $|\mu_k-\mu_q|\gg|\eta_{iq}\Omega_{ik}|,\,q\in\mathcal Q,\, q\neq k$, as well as taking a common spin phase $\phi_{ik}=\phi$ as assumed in the main text. For multiple simultaneous drives, the second-order expansion also contains terms involving two different drive frequencies. We assume that the drives are sufficiently spectrally separated that these $k\neq\ell$ contributions average to zero on the relevant evolution timescale ($|\mu_k-\mu_\ell|t\gg1,\,k\neq\ell$). The evolution operator generated by $H_{\rm cr}(t)$ is written using the Magnus expansion as
\begin{equation}
    U_{\rm cr}(t)=\exp\left[\overline{\Omega}_1(t)+\overline{\Omega}_2(t)\right],
\end{equation}
where
\begin{equation}
    \overline{\Omega}_1(t)=-i\int_0^t dt_1\,H_{\rm cr}(t_1),
\label{eq:supp_Omega1}
\end{equation}
and
\begin{equation}
    \overline{\Omega}_2(t)=-\frac{1}{2}\int_0^t dt_1\int_0^{t_1}dt_2\,[H_{\rm cr}(t_1),H_{\rm cr}(t_2)].
\label{eq:supp_Omega2}
\end{equation}
The first-order term describes residual spin-dependent phase-space displacements and can be written as
\begin{equation}
    \overline{\Omega}_1(t)=\sum_{i=1}^{N}\sum_{k=1}^{M}\sum_{m\notin\mathcal Q}\sigma_i^\phi\left[\alpha_{ikm}(t)a_m^\dagger-\alpha_{ikm}^*(t)a_m\right]+\sum_{i=1}^{N}\sum_{k=1}^{M}\sum_{\substack{q\in\mathcal Q\\q\neq k}}\sigma_i^\phi\left[\beta_{ikq}(t)a_q^\dagger-\beta_{ikq}^*(t)a_q\right],
\label{eq:supp_Omega1_explicit}
\end{equation}
where the phase-space displacements are given by
\begin{eqnarray}
    \alpha_{ikm}(t)&=-\frac{i\eta_{im}\Omega_{ik}}{\mu_k^2-\nu_m^2}\Bigg\{e^{i\nu_m t}\left[\mu_k\sin(\mu_k t+\psi_{ik})+i\nu_m\cos(\mu_k t+\psi_{ik})\right]-\mu_k\sin\psi_{ik}-i\nu_m\cos\psi_{ik}\Bigg\},\\
    \beta_{ikq}(t)&=-\frac{i\eta_{iq}\Omega_{ik}}{\mu_k^2-\mu_q^2}\Bigg\{e^{i\mu_q t}\left[\mu_k\sin(\mu_k t+\psi_{ik})+i\mu_q\cos(\mu_k t+\psi_{ik})\right]-\mu_k\sin\psi_{ik}-i\mu_q\cos\psi_{ik}\Bigg\}.
\end{eqnarray}
In the spectrally resolved regime considered here, these phase-space displacements remain bounded and perturbatively small.

For the second-order term, using the bosonic commutation relation $[a_p,a_q^\dagger]=\delta_{pq}$
and retaining only the same-drive contributions after the time averaging implied by $|\mu_k-\mu_\ell|t\gg1,\,k\neq\ell$, we obtain terms of the form
\begin{equation}
\begin{split}
    \overline{\Omega}_2(t)=&-\frac{1}{2}\sum_{i,j=1}^{N}\sum_{k=1}^{M}\sum_{m\notin\mathcal Q}\eta_{im}\eta_{jm}\Omega_{ik}\Omega_{jk}\sigma_i^\phi\sigma_j^\phi\int_0^t dt_1\int_0^{t_1}dt_2\,\left[e^{-i\nu_m(t_1-t_2)}-e^{i\nu_m(t_1-t_2)}\right]\\
    &\qquad\times\cos(\mu_k t_1+\psi_{ik})\cos(\mu_k t_2+\psi_{jk})\\
    &-\frac{1}{2}\sum_{i,j=1}^{N}\sum_{k=1}^{M}\sum_{\substack{q\in\mathcal Q\\q\neq k}}\eta_{iq}\eta_{jq}\Omega_{ik}\Omega_{jk}\sigma_i^\phi\sigma_j^\phi\int_0^t dt_1\int_0^{t_1}dt_2\,\left[e^{-i\mu_q(t_1-t_2)}-e^{i\mu_q(t_1-t_2)}\right]\\
    &\qquad\times\cos(\mu_k t_1+\psi_{ik})\cos(\mu_k t_2+\psi_{jk}).
\end{split}
\label{eq:supp_Omega2_explicit}
\end{equation}
The explicit integrals contain bounded oscillatory contributions as well as secular contributions proportional to $t$. On the timescale relevant to the quantum simulation, the bounded contributions do not accumulate, whereas the secular terms generate an effective spin--spin interaction. Retaining only these secular contributions gives
\begin{equation}
    \overline{\Omega}_2^{(\rm sec)}(t)\simeq-it\sum_{i<j}\sum_{k=1}^{M}\Omega_{ik}\Omega_{jk}\cos(\psi_{ik}-\psi_{jk})\sigma_i^\phi\sigma_j^\phi\left[\sum_{m\notin\mathcal Q}\frac{\eta_{im}\eta_{jm}\nu_m}{\mu_k^2-\nu_m^2}+\sum_{\substack{q\in\mathcal Q\\q\neq k}}\frac{\eta_{iq}\eta_{jq}\mu_q}{\mu_k^2-\mu_q^2}\right].
\label{eq:supp_Omega2_secular}
\end{equation}
In going from the double sum over $i$ and $j$ to $\sum_{i<j}$, the terms with $i=j$ have been omitted. Since $(\sigma_i^\phi)^2=\mathbb I$, these terms contribute only a spin-independent scalar phase and therefore do not affect the dynamics. Thus, the leading accumulated effect of the residual spin--motion couplings is described by
\begin{equation}
    H_{\rm cr}^{\rm eff}=\sum_{i<j}J_{ij}^{\rm cr}\sigma_i^\phi\sigma_j^\phi,
\label{eq:supp_Hcr_eff}
\end{equation}
where
\begin{equation}
    J_{ij}^{\rm cr}=\sum_{k=1}^{M}\Omega_{ik}\Omega_{jk}\cos(\psi_{ik}-\psi_{jk})\left[\sum_{m\notin\mathcal Q}\frac{\eta_{im}\eta_{jm}\nu_m}{\mu_k^2-\nu_m^2}+\sum_{\substack{q\in\mathcal Q\\q\neq k}}\frac{\eta_{iq}\eta_{jq}\mu_q}{\mu_k^2-\mu_q^2}\right].
\label{eq:supp_Jcr_eta}
\end{equation}
By defining the photon recoil frequency $\omega_{\rm rec}\equiv\Delta k^2/(2m_{\rm ion})$, we obtain
\begin{equation}
    J_{ij}^{\rm cr}=\omega_{\rm rec}\sum_{k=1}^{M}\Omega_{ik}\Omega_{jk}\cos(\psi_{ik}-\psi_{jk})\left[\sum_{m\notin\mathcal Q}\frac{\xi_{im}\xi_{jm}}{\mu_k^2-\nu_m^2}+\sum_{\substack{q\in\mathcal Q\\q\neq k}}\frac{\xi_{iq}\xi_{jq}}{\mu_k^2-\mu_q^2}\frac{\mu_q}{\nu_q}\right].
\label{eq:supp_Jcr}
\end{equation}

Combining the resonant spin--phonon interaction, the effective phonon Hamiltonian, and the residual spin--spin coupling gives
\begin{equation}
    H_{\rm eff}=\sum_{i=1}^{N}\sum_{k=1}^{M}\frac{g_{ik}}{2}\sigma_i^\phi\left(a_k e^{i\psi_{ik}}+a_k^\dagger e^{-i\psi_{ik}}\right)+\sum_{k=1}^{M}\omega_k a_k^\dagger a_k+\sum_{i<j}J_{ij}^{\rm cr}\sigma_i^\phi\sigma_j^\phi.
\label{eq:supp_final_general}
\end{equation}
In the main text, we choose $\psi_{ik}=0$ such that the spin couples to the position quadrature of each addressed mode. The Hamiltonian then reduces to
\begin{equation}
    H_{\rm eff}=\sum_{i=1}^{N}\sum_{k=1}^{M}\frac{g_{ik}}{2}\sigma_i^\phi(a_k+a_k^\dagger)+\sum_{k=1}^{M}\omega_k a_k^\dagger a_k+\sum_{i<j}J_{ij}^{\rm cr}\sigma_i^\phi\sigma_j^\phi.
\label{eq:supp_final_phi}
\end{equation}
Finally, working in the dressed-spin basis in which the common $\phi$ axis is mapped to the $z$ axis, $\sigma_i^\phi\mapsto\sigma_i^z$, gives
\begin{equation}
    H_{\rm eff}=\sum_{i=1}^{N}\sum_{k=1}^{M}\frac{g_{ik}}{2}\sigma_i^z(a_k+a_k^\dagger)+\sum_{k=1}^{M}\omega_k a_k^\dagger a_k+\sum_{i<j}J_{ij}^{\rm cr}\sigma_i^z\sigma_j^z.
\label{eq:supp_final_z}
\end{equation}
This is the effective spin--phonon Hamiltonian used in the main text, including the residual two-body spin interaction generated by the off-resonant coupling of each bichromatic drive to spectator modes and to addressed modes associated with the other drives. Within the single-excitation manifold, the residual spin--spin interaction reduces to a static energy shift~\cite{padilla2025vibrationallyassistedexcitontransfer}.

\end{document}